\documentclass[twocolumn,amsthm]{autart}

\usepackage{graphicx}
\usepackage{amsmath,amssymb,amsfonts}
\usepackage{graphicx}
\usepackage{mathtools}
\usepackage{mathrsfs}
\usepackage{tikz}
\usepackage{enumitem}

\newcommand{\bbR}{\mathbb{R}}
\newcommand{\bbN}{\mathbb{N}}
\newcommand{\bbE}{\mathbb{E}}

\newcommand{\bbQ}{\mathbb{Q}}
\newcommand{\bbP}{\mathbb{P}}

\newcommand{\calF}{\mathcal{F}}

\newcommand{\sfd}{\mathsf{d}}

\newcommand{\scrL}{\mathscr{L}}

\newcommand{\circnum}[1]{\textcircled{\raisebox{-0.15ex}{\scriptsize #1}}}

\newtheorem{theorem}{Theorem}
\newtheorem{lemma}[theorem]{Lemma}
\newtheorem{corollary}[theorem]{Corollary}
\newtheorem{proposition}[theorem]{Proposition}
\newtheorem{definition}{Definition}

\newtheorem{hypothesis}{Hypothesis}
\newtheorem{remark}{Remark}
\newtheorem{example}{Example}

\DeclareMathOperator*{\esssup}{ess\:sup}

\begin{document}

\begin{frontmatter}
\title{Continuous-Time Stochastic Contraction Induced by Common White Noise\thanksref{footnoteinfo}}%

\thanks[footnoteinfo]{This work was supported by the National Research Foundation of Korea (NRF) grant funded by the Korea government (MSIT) (RS-2026-25504174). Corresponding author: Jin Gyu Lee.}

\author[SNU]{Sunghyun Koo}\ead{shkoo@cdsl.kr},
\author[SNU]{Jin Gyu Lee}\ead{jingyu.lee@snu.ac.kr},
\author[SNU]{Hyungbo Shim}\ead{hshim@snu.ac.kr}

\address[SNU]{ASRI, Department of Electrical and Computer Engineering, Seoul National University, Seoul, Republic of Korea}


\begin{abstract}
This paper studies contraction properties of stochastic dynamical systems governed by It{\^o} stochastic differential equations.
We consider two trajectories driven by a common noise process and measure their separation using Finsler distances, which encompass Riemannian and norm-induced distances.
Two notions of contraction are then investigated: $L^p$-contraction, defined by exponential decay of the $L^p$-norm (or the $p$-th moment) of the distance, and almost sure contraction, defined by exponential decay of the distance along almost every sample path.
Each notion is characterized by a differential condition that is necessary and sufficient for contraction under mild hypotheses.
Relations between contraction notions and an associated incremental input-to-state stability result are also established.
As an illustration, we apply the established theory to analyze noise-induced synchronization and quantify its robustness with respect to parameter perturbations.
\end{abstract}

\end{frontmatter}

\section{Introduction}
\label{sec:introduction}

Contraction theory, the study of convergence of system trajectories toward one another, has become a powerful tool for analysis and control of nonlinear systems \cite{lohmiller98aut,sontag2010,FS14,aminzare2014cdc,tsukamoto2021arc}
since its introduction to the field by Lohmiller and Slotine \cite{lohmiller98aut}.
Because contraction ensures forgetting of initial conditions and limits error accumulation, many control and observer designs seek contractive closed-loop dynamics \cite{lohmiller2000tac,manchester2017tac,sasfi2023aut,sanfelice2012tac,bonnabel2015tac,yi2022tac}.
The same properties have also led to system analysis techniques, such as model reduction for singularly perturbed systems over an infinite time interval \cite{hoppensteadt,kim2015tac,lee2020aut}.
One can apply contraction theory to establish stability, synchronization, or entrainment without specifying an equilibrium point or a limit cycle \emph{a priori} \cite{manchester2014scl,russo2010plos,lee2022cdc,cecconi2026ejc}.

To extend such benefits to stochastic systems, recent efforts have been made to develop stochastic contraction theory.
Depending on applications, existing studies have considered different relationships between the noise processes driving two trajectories.
On the one hand, \cite{pham2009tac,dani2015tac,tsukamoto2021tac,han2022rnc} consider two trajectories driven by independent noise processes (including the degenerate case in which one trajectory is noise-free).
Assuming the noise-free system is contracting, these works establish an ultimate bound on the $L^2$-norm of the distance between the trajectories.\footnote{The $L^p$-norm of a real-valued random variable $X$ is defined by $\|X\|_{L^p} \coloneqq (\bbE|X|^p)^{1/p}$. This is a norm on a space of random variables, and it should be distinguished from norms on the state space $\bbR^n$.}
Such results have been used in robust controller and observer design.
Other studies, on the other hand, consider trajectories driven by a common noise process and derive sufficient conditions for their contraction, which turn out either to coincide with or to differ from the deterministic contraction conditions.
In particular, \cite{ahmad2010,amin22lcss,kawano24tac,kawano25tac} consider
contraction in the $L^p$-norm, \cite{monm23stoch,liu21tac} consider almost sure contraction, and \cite{ngoc24ijc} considers both notions.
The established conditions have been applied to analysis of noise-induced synchronization \cite{amin22lcss}, state estimation of an epidemic model \cite{kawano24tac}, and Wasserstein contraction for Langevin dynamics \cite{monm23stoch}.

The present paper aims to provide a more comprehensive study of common-noise-induced contraction, especially for It{\^o} stochastic differential equations (SDEs).
Except for the discrete-time results \cite{kawano24tac,kawano25tac}, existing works in this direction have used norm-induced distances to quantify the separation between trajectories.
In this paper, we consider time-dependent Finsler distances, which encompass both norm-induced and Riemannian distances \cite{FS14}.
We then derive differential conditions that guarantee each of $L^p$- and almost sure contraction in a Finsler distance, enabling local verification of contraction as in the deterministic case \cite{lohmiller98aut}.

We also address several aspects that have received limited attention in the existing literature.
First, we show that the proposed differential conditions are also necessary for contraction when the underlying Finsler structure is convex in each tangent space, as is the case for norm-induced and Riemannian distances.
Second, we clarify the relationships among various notions of stochastic contraction that have appeared in the literature.
Finally, we study incremental input-to-state stability in the present
common-noise setting, considering two trajectories driven
by the same noise process, one uncontrolled and the other subject to an input.

An interesting consequence of the established differential conditions is that a stochastic system can be contracting even when its noise-free counterpart is not.
This suggests a connection with noise-induced synchronization, a phenomenon observed across fields such as neuroscience \cite{neuro1,neuro2}, quantum systems \cite{quantum1,quantum2}, and ecology \cite{ecology}, in which non-contracting systems synchronize under a common stochastic forcing.
Motivated by this connection, we apply the proposed theory to analyze noise-induced synchronization in a double-well Langevin dynamics.
In particular, we construct a Finsler distance that certifies contraction and thereby establish noise-induced synchronization.
We also analyze the robustness of synchronization with respect to parameter perturbations using the incremental input-to-state stability result.
Complementing the results based on random attractors \cite{crauel1998,robinson2002}, our contraction-based analysis provides quantitative bounds on the synchronization probability.

The rest of the paper is organized as follows.
Section~\ref{sec:problem_setup} introduces the problem setup and several hypotheses used later in the paper.
Sections~\ref{sec:contraction_in_lp_norm} and \ref{sec:almost_sure_contraction} introduce the notions of $L^p$-contraction and almost sure contraction, respectively, and establish differential conditions characterizing these properties.
Section~\ref{sec:special_cases} specializes the established conditions to Riemannian and norm-induced distances.
Further discussions, including incremental input-to-state stability, are provided in Section~\ref{sec:further_discussions}.
Section~\ref{sec:application} analyzes noise-induced synchronization from a contraction perspective, and Section~\ref{sec:conclusion} concludes the paper.

\textit{Notation:}
We denote the underlying filtered probability space by $(\Omega,\mathcal{A},(\mathcal{F}_t)_{t\in\mathbb{R}},\mathbb{P})$. We assume that the usual conditions hold, that is, $(\mathcal{F}_t)_{t\in\mathbb{R}}$ is complete and right-continuous.
The $p$-norm of a vector is denoted by $\|\cdot\|_p$, the Frobenius norm of a matrix is denoted by $\|\cdot\|_F$, and the $L^p$-norm of a real-valued random variable is denoted by $\|\cdot\|_{L^p}$.
The abbreviation ``a.s.'' stands for ``almost surely.'' Both $\partial_x$ and $\frac{\partial}{\partial x}$ denote the partial derivative with respect to $x$.
For $a,b\in\bbR$, we denote $a\wedge b \coloneqq \min\{a,b\}$ and $a\vee b \coloneqq \max\{a,b\}$.

\section{Problem Setup}
\label{sec:problem_setup}

In this section, we introduce the system, its trajectories driven by a common noise process, Finsler distances, and contraction modes considered in this paper.
We also collect several hypotheses on the system and Finsler distances for later use.

\subsection{System and Trajectories}

In this paper, we consider a stochastic dynamical system governed by an It{\^o} SDE
\begin{equation}
\label{eqn:sde}
    dx = f(t,x)dt + \sigma(t,x) dW_t,
\end{equation}
where $x\in\bbR^n$ denotes the state, $f:\bbR\times \bbR^n \rightarrow \bbR^n$ and $\sigma:\bbR\times \bbR^n \rightarrow \bbR^{n\times m}$ are Borel measurable functions, and $W_t$ is an $m$-dimensional Wiener process (or Brownian motion) with respect to the underlying filtration $(\calF_t)_{t\in\bbR}$.
Throughout the paper, we assume that \eqref{eqn:sde} admits a unique maximal solution $(\phi_{t_0,t}(\xi))_{\tau_\infty(t_0,\xi) > t\geq t_0}$ for every initial time $t_0\in\bbR$ and initial condition $\xi\in\bbR^n$.\footnote{For example, Hypothesis~\ref{hyp:regularity}-(a) and (b) guarantee this.} Here, $\tau_\infty(t_0,\xi):\Omega\rightarrow (t_0,\infty]$ denotes the explosion time.
In addition, we assume that the solution is \emph{strongly complete}, that is, $\tau_{\infty}(t_0,\xi)=\infty$ for all $\xi\in\bbR^n$ almost surely for each $t_0\in\bbR$ (cf. \cite{li2011prob}). Sufficient conditions for strong completeness can be found in, e.g., \cite{scheutzow17}.

To study contraction, we focus on a pair of trajectories of \eqref{eqn:sde} with possibly different initial conditions, both driven by a common noise process.
We denote such trajectories by $x_1(t)=\phi_{t_0,t}(\xi_1)$ and $x_2(t)=\phi_{t_0,t}(\xi_2)$, which solve the $2n$-dimensional SDE
\begin{equation*}
    \begin{bmatrix}
        dx_1\\dx_2
    \end{bmatrix} = \begin{bmatrix}
        f(t,x_1)\\f(t,x_2)
    \end{bmatrix}dt + \begin{bmatrix}
        \sigma(t,x_1)\\\sigma(t,x_2)
    \end{bmatrix}dW_t
\end{equation*}
with initial conditions $x_1(t_0)=\xi_1$ and $x_2(t_0)=\xi_2$.

We often impose the following regularity on \eqref{eqn:sde}.
Let $\sigma_j(t,x)$ denote the $j$-th column of $\sigma(t,x)$.

\begin{hypothesis}
\label{hyp:regularity}
    System \eqref{eqn:sde} satisfies the following:
    \begin{enumerate}[label=(\alph*),nosep]
        \item $f$ and $\sigma$ are right-continuous with left limits (i.e., c{\`a}dl{\`a}g) in $t$ for each $x\in\bbR^n$,
        \item
        \label{item:local_Lipschitz_uniformly_in_time}
        $f$ and $\sigma$ are locally Lipschitz in $x$ locally uniformly in $t$,
        \item $f$ and $\sigma$ are differentiable in $x$ for each $t\in\bbR$,
        \item
        \label{item:local_Holder_uniformly_in_time}
        $\partial_x f$ and each $\partial_x\sigma_j$ are locally $\alpha$-H{\"o}lder continuous in $x$ locally uniformly in $t$ for some $\alpha>0$.
    \end{enumerate}
\end{hypothesis}

Formal statements of conditions (b) and (d) can be found in Appendix~\ref{appendix:regularity_details}.
We note that Hypothesis~\ref{hyp:regularity} is satisfied when $f(t,x)$ and $\sigma(t,x)$ are $C^2$.

Under Hypothesis~\ref{hyp:regularity}, the maximal solutions of 
\eqref{eqn:sde} constitute a forward stochastic flow of local $C^1$-diffeomorphisms \cite[Thm.~4.7.2]{kunita}.
The following lemma is a direct consequence of this fact and forms the basis of our differential analysis.
Let $W_t^j$ denote the $j$-th component of $W_t$.

\begin{lemma}
\label{lem:stochastic_flow}
    Suppose that Hypothesis~\ref{hyp:regularity} holds.
    Then, for the unique maximal solution $(\phi_{t_0,t}(\xi))_{t\geq t_0}$ of \eqref{eqn:sde},
    \begin{enumerate}[label=(\alph*),nosep]
        \item $\phi_{t_0,t}(\xi)$ is continuous in $(t_0,t,\xi)$,
        \item $\phi_{t_0,t}(\xi)$ is differentiable in $\xi$,
        \item $\partial_\xi \phi_{t_0,t}(\xi)$ is continuous in $(t_0,t,\xi)$,
        \item $\partial_\xi \phi_{t_0,t}(\xi)$ is nonsingular
    \end{enumerate}
    for all $t\geq t_0$ and $\xi\in\bbR^n$ almost surely.
    Furthermore, $(\partial_\xi \phi_{t_0,t}(\xi))_{t\geq t_0}$ satisfies
    \begin{equation}
    \label{eqn:matrix_sde}
    \begin{aligned}
        \frac{\partial\phi_{t_0,t}}{\partial \xi}(\xi) = I + \int_{t_0}^t \frac{\partial f}{\partial x}(s,\phi_{t_0,s}(\xi)) \frac{\partial\phi_{t_0,s}}{\partial \xi}(\xi) ds&\\
        +\sum_{j=1}^m\int_{t_0}^t \frac{\partial \sigma_j}{\partial x}(s,\phi_{t_0,s}(\xi)) \frac{\partial\phi_{t_0,s}}{\partial \xi}(\xi) dW_s^j&
    \end{aligned}
    \end{equation}
    for all $t\geq t_0$ and $\xi\in\bbR^n$.
\end{lemma}

\begin{proof}
    Given the strong completeness of \eqref{eqn:sde}, (a)--(c) follow from \cite[Thm.~4.7.2]{kunita}. Item~(d) and \eqref{eqn:matrix_sde} appear in the proof of \cite[Thm.~4.6.5]{kunita}. See also \cite[Thm.~3.4 and 3.11]{kunita2}.
\end{proof}

\subsection{Finsler Distance}

A key ingredient in contraction analysis is a suitable, possibly time-dependent distance function $\sfd_t(\cdot,\cdot)$ that quantifies the separation between two trajectories.
The most commonly used choices are Riemannian \cite{lohmiller98aut} and norm-induced distances \cite{sontag2010}, and Finsler distances form a broader class that encompasses both as special cases \cite{FS14}.
In this respect, we work with general time-dependent Finsler distances.

We first introduce (time-dependent) Finsler structures.
Let $T\bbR^n$ denote the tangent bundle of $\bbR^n$, which can be identified with $\bbR^n\times \bbR^n$.
Let an element $(x,v)\in T \bbR^n$ represent the point $x\in\bbR^n$ together with a tangent vector $v\in\bbR^n$.

\begin{definition}
    \label{def:finsler_structure}
    A \emph{Finsler structure} is a family of continuous functions $F_t:T\bbR^n\rightarrow \bbR$, indexed by $t\in\bbR$, such that
    \begin{enumerate}[label=(\alph*),nosep]
        \item (positive homogeneity) $F_t(x,\lambda v)=\lambda F_t(x,v)$ for all $\lambda \geq 0$ for each $(x,v)\in T\bbR^n$, 
        \item (positive definiteness) $F_t(x,v)>0$ whenever $v\neq0$.
    \end{enumerate}
\end{definition}

A Finsler structure naturally defines the notion of curve length for each $t\in\bbR$.
Specifically, for each piecewise $C^1$ curve $\gamma:[a,b]\rightarrow \bbR^n$, we can define its length as
\begin{equation}
\label{eqn:curve_length}
    \ell_t(\gamma) \coloneqq \int_a^b F_t(\gamma(r),\dot\gamma(r))dr,
\end{equation}
understanding $F_t(\gamma(r),\dot\gamma(r))$ as the cost of moving from $\gamma(r)$ in the direction $\dot\gamma(r)$.
Indeed, $\ell_t(\gamma)$ is invariant under orientation-preserving reparameterizations.

This length structure further defines an asymmetric distance function on $\bbR^n$, which is called a Finsler distance \cite[Example~2.2.5]{BBI01}.

\begin{definition}
    The \emph{Finsler distance} induced by a Finsler structure $F_t$ is defined by
    \begin{equation*}
        \sfd_t(x,x') \coloneqq \inf_{\gamma\in\Gamma(x,x')} \ell_t(\gamma)
    \end{equation*}
    for $x,x'\in\bbR^n$, where $\Gamma(x,x')$ denotes the collection of all regular piecewise $C^1$ curves\footnote{A regular piecewise $C^1$ curve is a parameterized curve that is $C^1$ and has nonzero derivative on each subinterval of a finite partition of its domain.} that connect $x$ to $x'$, and $\ell_t(\gamma)$ denotes the length of $\gamma$, defined by \eqref{eqn:curve_length}.
\end{definition}

The following examples show how Riemannian and norm-induced distances fit into this Finsler-based framework.

\begin{example}[Riemannian distance]
\label{ex:riemannian}
    A Riemannian distance is a distance function induced by a Finsler structure of the form $F_t(x,v)=\sqrt{v^\top M_t(x)v}$ where $M_t(x)\in\bbR^{n\times n}$ is symmetric and positive definite for every $(t,x)\in\bbR\times\bbR^n$.
\end{example}

\begin{example}[Norm-induced distance]
\label{ex:norm}
    A norm $|\cdot|$ defined on $\bbR^n$ induces a distance function $\sfd(x_1,x_2)=|x_1-x_2|$ on $\bbR^n$. This is precisely the distance function induced by the Finsler structure $F_t(x,v)=|v|$.
\end{example}

In the following, we introduce some hypotheses on Finsler structures.
Let $T\bbR^n \setminus 0 \coloneqq \{(x,v)\in T\bbR^n : v\neq0\}$ denote the slit tangent bundle of $\bbR^n$.

\begin{hypothesis}
\label{hyp:smoothness}
    The partial derivatives
    \begin{equation*}
        \frac{\partial F_t}{\partial t}, \frac{\partial F_t}{\partial x},\frac{\partial F_t}{\partial v},\frac{\partial^2 F_t}{\partial x^2},\frac{\partial^2 F_t}{\partial x\partial v},\frac{\partial^2 F_t}{\partial v^2}
    \end{equation*}
    exist as continuous functions in $(t,x,v)$ on $\bbR\times (T\bbR^n\setminus 0)$.
\end{hypothesis}

\begin{hypothesis}
\label{hyp:convexity}
    Finsler structure $F_t$ is convex in each tangent space, i.e., $F_t(x,\cdot)$ is convex for each $(t,x)\in\bbR\times\bbR^n$.
\end{hypothesis}

The role of Hypothesis~\ref{hyp:smoothness} is to ensure the applicability of It{\^o}'s formula.
Hypothesis~\ref{hyp:convexity} guarantees that the distance between two nearby points can be approximated using the Finsler structure, i.e.,
\begin{equation}
\label{eqn:distance_approximation}
    \sfd_t(x,x+v) \approx F_t(x,v)
\end{equation}
for $v\approx 0$ \cite[Thm.~4.3]{busemann41ams}.
Note that Hypothesis~\ref{hyp:convexity} is always satisfied by Riemannian and norm-induced distances.

\begin{remark}
    In some literature, e.g., \cite[Ch.~1]{BCS00}, a Finsler structure additionally requires smoothness of $F(\cdot,\cdot)$ and strong convexity of $F^2(x,\cdot)$.
    These conditions are central to the differential geometry of Finsler manifolds, including the theory of connections and curvature.
    In this paper, however, our objective is to define a sufficiently general notion of distance for contraction analysis.
    For this purpose, the weaker notion of a Finsler structure adopted here suffices.
\end{remark}

\subsection{Modes of Contraction}

Given two trajectories $x_1$ and $x_2$ and a Finsler distance $\sfd_t(\cdot,\cdot)$,
contraction theory asks whether $\sfd_t(x_1(t),x_2(t))$ decays exponentially or not.
Unlike in the deterministic setting, however, $\sfd_t(x_1(t),x_2(t))$ itself is now a random variable.
We thus need to specify in what sense this distance decays.

The authors of \cite{amin22lcss,kawano24tac,kawano25tac,ngoc24ijc} quantify the ``size" of the random distance $\sfd_t(x_1(t),x_2(t))$ using its $L^p$-norm
\begin{equation*}
    \|\sfd_t(x_1(t),x_2(t))\|_{L^p}
    \coloneqq \big(
        \bbE[\sfd_t^p(x_1(t),x_2(t))]
    \big)^{1/p},
\end{equation*}
and define contraction by the exponential decay of this norm.\footnote{Strictly speaking, the previous works have used the $p$-th moment of the distance rather than its $L^p$-norm.
Although equivalent, we insist on describing contraction in terms of the $L^p$-norm, as our analysis makes use of norm properties, such as the triangle inequality.}
Alternatively, \cite{monm23stoch,liu21tac,ngoc24ijc} characterize contraction by the decay of the distance along almost every sample path.
We shall refer to these two modes of contraction as $L^p$-contraction and almost sure contraction, respectively, and develop differential conditions that characterize each of them.

\section{${L}^{p}$-Contraction}
\label{sec:contraction_in_lp_norm}

\subsection{The Notion of ${L}^{p}$-contraction}

Extending the definition in \cite{amin22lcss} to general Finsler distances, we define $L^p$-contraction as follows.

\begin{definition}
\label{def:Lp_contraction}
For $p\in[1,\infty)$, stochastic dynamical system \eqref{eqn:sde} is said to be \emph{$L^p$-contracting} in distance $\sfd_t$ with rate $\lambda>0$ if
\begin{equation}
    \label{eqn:Lp_contraction}
        \|\sfd_t(x_1(t),x_2(t))\|_{L^p}\leq e^{-\lambda(t-t_0)}\sfd_{t_0}(\xi_1,\xi_2)\quad \forall t\geq t_0
\end{equation}
for every $t_0\in\bbR$ and $x_i(t_0)=\xi_i\in\bbR^n$, $i=1,2$.
\end{definition}

The following linear SDEs provide simple examples of $L^p$-contracting systems.

\begin{example}
\label{ex:ou_process}
    Consider an Ornstein-Uhlenbeck (OU) process
    \begin{equation}
    \label{eqn:ou_process}
        dx = \alpha x dt + \beta dW_t \in \bbR.
    \end{equation}
    Let $x_1$ and $x_2$ be two trajectories driven by a common noise process.
    If we define $e \coloneqq x_2-x_1$, we have $\dot e = \alpha e$.
    Thus, for each $p\in[1,\infty)$, the OU process \eqref{eqn:ou_process} is $L^p$-contracting in the Euclidean distance if and only if $\alpha<0$.
\end{example}

\begin{example}
\label{ex:geometric_brownian_motion}
    Consider a geometric Brownian motion
    \begin{equation}
    \label{eqn:geometric_brownian_motion}
        dx = \alpha x dt + \beta x dW_t \in \bbR.
    \end{equation}
    Let $x_1$ and $x_2$ be two trajectories driven by a common noise process.
    Since $e \coloneqq x_2-x_1$ is also a solution of \eqref{eqn:geometric_brownian_motion}, for each $p\in (0,\infty)$, we have \cite[Example~3.5.5]{Mao08}
    \begin{equation*}
        \bbE|e(t)|^p = e^{p(\alpha+{(p-1)\beta^2}/{2})(t-t_0)}|e(t_0)|^p.
    \end{equation*}
    Thus, the geometric Brownian motion \eqref{eqn:geometric_brownian_motion} is $L^p$-contracting in the Euclidean distance if and only if $\alpha < -\frac{(p-1)\beta^2}{2}$.
\end{example}

Example~\ref{ex:geometric_brownian_motion} illustrates that $L^p$-contraction becomes more stringent as $p$ increases. This observation holds in general due to the monotonicity of $L^p$-norms \cite[Sec.~6.7]{Will91}.

\begin{proposition}
\label{prop:Lp_hierarchy}
    If system~\eqref{eqn:sde} is $L^p$-contracting for some $p\in[1,\infty)$, then it is $L^q$-contracting for every $q\in[1,p]$ in the same distance with the same rate.
\end{proposition}

\begin{proof}
    For each $q\in[1,p]$, we have $\|\sfd_t(x_1(t),x_2(t))\|_{L^q} \leq \|\sfd_t(x_1(t),x_2(t))\|_{L^p} \leq e^{-\lambda(t-t_0)} \sfd_{t_0}(\xi_1,\xi_2)$.
\end{proof}

Definition~\ref{def:Lp_contraction} is stated only for trajectories starting from deterministic initial conditions.
Nevertheless, the contraction property extends automatically to random initial conditions, as noted in \cite[Remark~2.2]{kawano24tac}, and to arbitrary time intervals.

\begin{proposition}
\label{prop:Lp_random_initial_condition}
    Let $t_0\in\bbR$ and $p\in[1,\infty)$. Let $\xi_1$ and $\xi_2$ be $\bbR^n$-valued $\calF_{t_0}$-measurable random variables.
    Let $x_1(t_0)=\xi_1$ and $x_2(t_0)=\xi_2$.
    If system \eqref{eqn:sde} is $L^p$-contracting in distance~$\sfd_t$ with rate $\lambda>0$, then
    \begin{equation*}
        \| \sfd_{t}(x_1(t),x_2(t))\|_{L^p} \leq e^{-\lambda(t-s)} \| \sfd_{s}(x_1(s),x_2(s))\|_{L^p}
    \end{equation*}
    for any $t\geq s\geq t_0$.
\end{proposition}

\begin{proof}
    Conditioning on $\calF_{s}$ (cf. \cite[Lemma~A.3]{schilling}), we have $\bbE[\sfd_{t}^p(x_1(t),x_2(t))|\calF_{s}] \leq e^{-p\lambda(t-s)} \sfd_{s}^p(x_1(s),x_2(s))$ almost surely.
    The result then follows from the tower property.
\end{proof}

\subsection{Differential Characterization}
\label{sec:Lp_differential_characterization}

We now derive a differential condition that characterizes $L^p$-contraction.
As a first step, motivated by \cite{lohmiller98aut,FS14}, we introduce the notion of variational (or virtual) dynamics.
Let $x(t)$ and $x(t)+v(t)$ be two nearby trajectories of system~\eqref{eqn:sde}, where $v(t)\approx 0$ denotes an infinitesimal displacement between the two trajectories.
Then, one would expect $v$ to evolve according to the variational dynamics
\begin{equation}
\label{eqn:variational_sde}
    dv = \frac{\partial f}{\partial x}(t,x)v dt + \sum_{j=1}^m \frac{\partial \sigma_j}{\partial x}(t,x)v dW_t^j,
\end{equation}
which is indeed considered in \cite{liu21tac}.
Since the parameters of~\eqref{eqn:variational_sde} are functions of $x$, it can be more precisely viewed as a part of the $2n$-dimensional SDE
\begin{equation}
\label{eqn:lifted_sde}
    \begin{bmatrix}
        dx\\dv
    \end{bmatrix} = \begin{bmatrix}
        f(t,x)\\
        \frac{\partial f}{\partial x}(t,x)v
    \end{bmatrix}dt + \sum_{j=1}^m\begin{bmatrix}
        \sigma_j(t,x)\\
        \frac{\partial \sigma_j}{\partial x}(t,x)v
    \end{bmatrix}dW_t^j.
\end{equation}

In view of \eqref{eqn:distance_approximation}, to establish the $L^p$-contraction property for these two nearby trajectories, we are led to verify
\begin{equation}
\label{eqn:Lp_infinitesimal}
    \bbE[F_t^p(x(t),v(t))]\leq e^{-p\lambda (t-t_0)} F_{t_0}^p(\xi,\zeta),
\end{equation}
where $\xi,\zeta\in\bbR^n$ denote the initial conditions of $x(t)$ and $v(t)$, respectively.
Meanwhile, a heuristic application of Dynkin's formula \cite[p.~10]{kushner} suggests
\begin{equation*}
    \bbE[F_t^p(x(t),v(t))] = F_{t_0}^p(\xi,\zeta) + \int_{t_0}^t \bbE[\scrL F_s^p(x(s),v(s))]ds,
\end{equation*}
where $\scrL$ denotes the differential generator of \eqref{eqn:lifted_sde}, which in fact coincides with its infinitesimal generator \cite[pp.~15--16]{kushner}.
Therefore, we may conclude that the differential inequality
\begin{equation*}
    \scrL F_t^p(x,v) \leq -p\lambda F_t^p(x,v)
\end{equation*}
should imply \eqref{eqn:Lp_infinitesimal}.
Expanding the left-hand side, we obtain the differential condition
\begin{equation}
\label{eqn:finsler_condition}
    \frac{\partial F_t^p}{\partial t} + \frac{\partial F_t^p}{\partial x} f + \frac{\partial F_t^p}{\partial v} \frac{\partial f}{\partial x} v + \sum_{j=1}^m \Delta_{j,p} \leq -p\lambda {F_t^p},
\end{equation}
where $\Delta_{j,p}(t,x,v)$ is defined by
\begin{align*}
    \Delta_{j,p}(t,x,v) \coloneqq \frac{1}{2} \sigma_j^\top \left[\frac{\partial^2 F_t^p}{\partial x^2}\right] \sigma_j + \sigma_j^\top\left[\frac{\partial^2 F_t^p}{\partial x\partial v}\right]\left( \frac{\partial \sigma_j}{\partial x}v\right)&\\
    + \frac{1}{2}\left(\frac{\partial \sigma_j}{\partial x} v\right)^\top \left[\frac{\partial^2 F_t^p}{\partial v^2}\right] \left(\frac{\partial \sigma_j}{\partial x}v\right)&.
\end{align*}
Note that all the partial derivatives appearing in \eqref{eqn:finsler_condition} are well-defined on $\bbR\times (T\bbR^n\setminus0)$ under Hypothesis~\ref{hyp:smoothness}.

The derivation so far identifies \eqref{eqn:finsler_condition} as a candidate sufficient condition for $L^p$-contraction, but some gaps remain to be filled in a mathematically rigorous manner.
Indeed, we prove that \eqref{eqn:finsler_condition} is sufficient for $L^p$-contraction under suitable hypotheses.
We also show that \eqref{eqn:finsler_condition} is necessary for $L^p$-contraction under an additional hypothesis.

\begin{theorem}
\label{thm:Lp_contraction}
    Let $F_t(\cdot,\cdot)$ be a Finsler structure and $\sfd_t(\cdot,\cdot)$ denote the induced Finsler distance.
    Suppose that Hypotheses~\ref{hyp:regularity} and~\ref{hyp:smoothness} hold.
    Let $p\in[1,\infty)$ and $\lambda>0$.
    \begin{enumerate}[label=(\alph*),nosep]
        \item If condition~\eqref{eqn:finsler_condition} holds for all $(t,x,v)\in\bbR\times (T\bbR^n\setminus0)$, then system~\eqref{eqn:sde} is $L^p$-contracting in $\sfd_t$ with rate $\lambda$.
        \item If system \eqref{eqn:sde} is $L^p$-contracting in $\sfd_t$ with rate $\lambda$, and the Finsler structure $F_t$ satisfies Hypothesis~\ref{hyp:convexity}, then condition~\eqref{eqn:finsler_condition} holds for all $(t,x,v)\in\bbR\times (T\bbR^n\setminus0)$.
    \end{enumerate}
\end{theorem}

\begin{proof}
    See Section~\ref{sec:proof_Lp}.
\end{proof}

Before moving on to the formal proof, we discuss several aspects of the differential condition~\eqref{eqn:finsler_condition}.

\begin{enumerate}[label=(\roman*),nosep]
    \item
    \label{item:comparison_with_deterministic_contraction}
    Condition~\eqref{eqn:finsler_condition} generalizes a contraction condition for deterministic systems.
    Indeed, when $\sigma_j=0$, the noise-related terms $\Delta_{j,p}$ vanish, and \eqref{eqn:finsler_condition} reduces to
    \begin{equation*}
        \frac{\partial F_t^p}{\partial t} + \frac{\partial F_t^p}{\partial x} f + \frac{\partial F_t^p}{\partial v} \frac{\partial f}{\partial x}v \leq -p\lambda F_t^p,
    \end{equation*}
    which coincides with the deterministic contraction criterion provided in \cite{FS14} except for some technical details.
    
    \item
    \label{item:noise_induced_contraction}
    Interestingly, since $\Delta_{j,p}$ can be negative, system~\eqref{eqn:sde} can be $L^p$-contracting even when the noise-free system $\dot x = f(t,x)$ is not contracting.
    We provide such an example in Section~\ref{sec:application}.
    In view of \cite{lee2022cdc}, this can be interpreted as a contraction induced by an open-loop white-noise input.
    
    \item
    \label{item:another_representation}
    In view of Proposition~\ref{prop:Lp_hierarchy}, condition~\eqref{eqn:finsler_condition} should be more stringent for a larger value of $p$.
    Indeed, for any $q > 0$, one can rewrite \eqref{eqn:finsler_condition} equivalently as
    \begin{equation}
    \label{eqn:another_finsler_condition}
        \frac{\partial F_t^q}{\partial t} + \frac{\partial F_t^q}{\partial x} f + \frac{\partial F_t^q}{\partial v} \frac{\partial f}{\partial x} v + \sum_{j=1}^m (\Delta_{j,q}+\tilde\Delta_{j,p,q}) \leq -q\lambda {F_t^q}
    \end{equation}
    using the chain rule. Here, $\tilde\Delta_{j,p,q}(t,x,v)$ is defined by
    \begin{align*}
        \tilde\Delta_{j,p,q}(t,x,v) \coloneqq \frac{p-q}{2qF_t^q}\left(\frac{\partial F_t^q}{\partial x}\sigma_j + \frac{\partial F_t^q}{\partial v}\frac{\partial \sigma_j}{\partial x}v\right)^2
    \end{align*}
    for all $(t,x,v)\in\bbR\times (T\bbR^n\setminus 0)$. (See Appendix~\ref{appendix:change_of_exponent} for its derivation.)
    It can now be seen that condition \eqref{eqn:finsler_condition} with $p$ replaced by $q\in[1,p]$ is no more stringent than the original condition because $\tilde\Delta_{j,p,q}$ is nonnegative.

    \item Given a Finsler structure $F_t$, there often exists a particular exponent $q$ for which the computation of $F_t^q$ is simpler, e.g., Riemannian cases with $q=2$ and Finsler structures induced by the $q$-norm.
    In such cases, condition~\eqref{eqn:another_finsler_condition} can be more convenient for verifying the contraction property.

    \item 
    \label{item:unit_tangent_bundle}
    Due to the positive homogeneity, a Finsler structure can be completely characterized by its restriction to the unit tangent bundle $UT\bbR^n \coloneqq \{(x,v)\in T\bbR^n:\|v\|_2 = 1\}$.
    Consistent with this, since all the terms in \eqref{eqn:finsler_condition} are positively homogeneous of degree $p$ in $v$, it suffices to verify \eqref{eqn:finsler_condition} only for all $(t,x,v)\in\bbR\times UT\bbR^n$.
    Therefore, $L^p$-contraction can be characterized using only information on the unit tangent bundle.

    \item 
    \label{item:existence}
    Condition \eqref{eqn:finsler_condition} does not guarantee the strong completeness of \eqref{eqn:sde}, or even the existence of a global solution.
    These properties are assumed throughout the paper and should be verified separately.

    \item
    \label{item:locally_bounded}
    For ease of presentation, Hypothesis~\ref{hyp:regularity} is stated in a form that is slightly more restrictive than the proof of Theorem~\ref{thm:Lp_contraction} actually requires.
    For Theorem~\ref{thm:Lp_contraction}-(a), Hypothesis~\ref{hyp:regularity}-(a) can be weakened to require that $f$ and $\sigma$ are locally bounded in $t$ for each $x\in\bbR^n$.
    For Theorem~\ref{thm:Lp_contraction}-(b), it suffices to additionally assume that $f$ and $\sigma$ are right-continuous in $t$ for each $x\in\bbR^n$.
    
\end{enumerate}

\subsection{Proof of Theorem~\ref{thm:Lp_contraction}}
\label{sec:proof_Lp}

We now turn to the proof of Theorem~\ref{thm:Lp_contraction}.

\subsubsection*{Sufficiency}
The proof of Theorem~\ref{thm:Lp_contraction}-(a) proceeds analogously to the construction of a Finsler distance:
\begin{itemize}[nosep]
    \item \textit{Step 1:} Contraction of infinitesimal displacements
    \item \textit{Step 2:} Contraction of curve lengths
    \item \textit{Step 3:} Contraction in the Finsler distance
\end{itemize}

\textit{Step 1:} We prove the infinitesimal contraction property \eqref{eqn:Lp_infinitesimal}.
Fix $(\xi,\zeta)\in T\bbR^n\setminus0$. Let $x(t)$ and $v(t)$ denote the solution of \eqref{eqn:lifted_sde} with initial conditions $x(t_0)=\xi$ and $v(t_0)=\zeta$.
Define $V(t,x,v)\coloneqq e^{p\lambda t}F_t^p(x,v)$. Then, it suffices to show that $\bbE[V(t,x(t),v(t))]\leq V(t_0,\xi,\zeta)$.

To this end, we introduce the stopping time $\tau_R \coloneqq \inf\{t\geq t_0: \|x(t)\| \vee \|v(t)\| \geq R\}$
for each $R>0$.
Then, by applying It{\^o}'s formula,\footnote{By Lemma~\ref{lem:v_nonzero} in Appendix~\ref{appendix:technical_lemmas}, we have $v(t)\neq 0$ for all $t\geq t_0$ almost surely. Thus, we can apply It{\^o}'s formula under Hypothesis~\ref{hyp:smoothness}.} we obtain
\begin{equation}
\label{eqn:stopped_ito_formula}
    \begin{aligned}
        &V(t\wedge \tau_R) = V(t_0) + \int_{t_0}^{t\wedge \tau_R} \scrL V(s) ds\\
        &\qquad+\sum_{j=1}^m\int_{t_0}^{t\wedge \tau_R} \left(\frac{\partial V}{\partial x}\sigma_j + \frac{\partial V}{\partial v}\frac{\partial \sigma_j}{\partial x}v\right)dW_s^j,
    \end{aligned}
\end{equation}
where $V(s)$ is shorthand for $V(s,x(s),v(s))$. Observe that
\begin{equation*}
    \scrL V(s,x,v) = p\lambda e^{p\lambda s} F_s^p(x,v) + e^{p\lambda s} \scrL F_s^p(x,v) \leq 0 
\end{equation*}
by \eqref{eqn:finsler_condition} and that the stochastic integral on the right-hand side of \eqref{eqn:stopped_ito_formula} is a true martingale. 
Thus, by taking the expectation on both sides of \eqref{eqn:stopped_ito_formula}, we have
\begin{equation*}
    \bbE[V(t\wedge \tau_R)] \leq V(t_0).
\end{equation*}
Taking $R\rightarrow \infty$ and applying Fatou's lemma now yields
\begin{equation*}
    \bbE[V(t)] \leq \liminf_{R\rightarrow\infty} \bbE[V(t\wedge \tau_R)] \leq V(t_0).
\end{equation*}
Note that $t\wedge \tau_R \rightarrow t$ due to the existence of a global solution, and that $V(t\wedge \tau_R)\rightarrow V(t)$ because $V$ is jointly continuous in $(t,x,v)$ under Hypothesis~\ref{hyp:smoothness}.

\textit{Step 2:} Let $\gamma$ be a regular piecewise $C^1$ curve.
Then, $\phi_{t_0,t}\circ \gamma$ represents the evolution of $\gamma$ along the dynamics \eqref{eqn:sde}.
We next prove that
\begin{equation*}
    \bbE[\ell_t^p(\phi_{t_0,t}\circ\gamma)] \leq e^{-p\lambda(t-t_0)} \ell_{t_0}^p(\gamma),
\end{equation*}
where $\ell_t(\cdot)$ denotes the curve length induced by $F_t$.
Note that $\ell_t(\phi_{t_0,t}\circ \gamma)$ is well-defined since $\phi_{t_0,t}\circ \gamma$ is also a regular piecewise $C^1$ curve by Lemma~\ref{lem:stochastic_flow}.
Denoting $L\coloneqq \ell_{t_0}(\gamma)$, we may choose a parametrization $\gamma:[0,L]\rightarrow\bbR^n$ such that $F_{t_0}(\gamma(r),\dot\gamma(r))=1$ for almost every $r\in[0,L]$.

\begin{figure}[t]
    \centering
    \includegraphics[width=0.75\columnwidth]{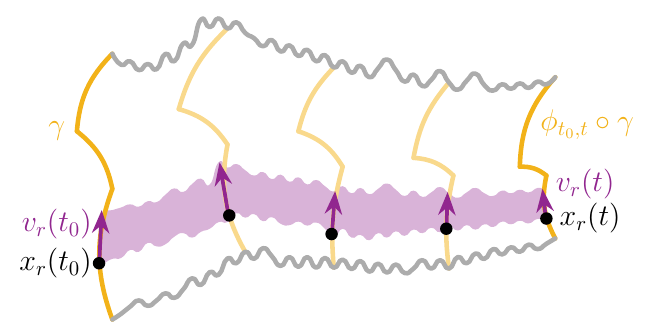}
    \caption{An illustration of the evolution of $x_r(t)$ and $v_r(t)$}
    \label{fig:flow_diagram}
\end{figure}

As illustrated in Fig.~\ref{fig:flow_diagram}, we define
\begin{equation}
\label{eqn:flow_of_curve}
    x_r(t) \coloneqq \phi_{t_0,t}(\gamma(r)),\quad v_r(t) \coloneqq \frac{\partial \phi_{t_0,t}(\gamma(r))}{\partial \xi} \dot\gamma(r)
\end{equation}
for all $t \geq t_0$ for each $r\in[0,L]$.
Then, we have
\begin{align*}
    \ell_{t}(\phi_{t_0,t}\circ \gamma) &= \int_0^L F_t(x_r(t),v_r(t))dr\\
    & \leq L^{(p-1)/p} \left(\int_0^L F_t^p(x_r(t),v_r(t)) dr\right)^{1/p}
\end{align*}
by H{\"o}lder's inequality.
Since $v_r(t_0)=\dot\gamma(r) \neq 0$ almost everywhere, by the result of Step 1, we obtain
\begin{align*}
    \bbE[\ell_t^p(\phi_{t_0,t}\circ \gamma)]
    &\leq L^{p-1} \int_0^L \bbE\left[F_t^p\left(x_r(t), v_r(t)\right)\right]dr\\
    &\leq L^{p-1} \int_0^L e^{-p\lambda(t-t_0)} F_{t_0}^p(\gamma(r),\dot\gamma(r))dr\\
    &= e^{-p\lambda(t-t_0)} L^p.
\end{align*}

\text{Step 3:} We now prove that system~\eqref{eqn:sde} is $L^p$-contracting.
Fix $\xi_1,\xi_2\in\bbR^n$ and $\gamma\in\Gamma(\xi_1,\xi_2)$.
Then, since $\phi_{t_0,t}\circ \gamma\in\Gamma(\phi_{t_0,t}(\xi_1),\phi_{t_0,t}(\xi_2))$, we have $\sfd_t(\phi_{t_0,t}(\xi_1),\phi_{t_0,t}(\xi_2)) \leq \ell_t(\phi_{t_0,t}\circ \gamma)$.
Applying the result of Step 2, we obtain
\begin{align*}
    \bbE[\sfd_t^p(\phi_{t_0,t}(\xi_1),\phi_{t_0,t}(\xi_2))] &\leq \bbE[\ell_t^p(\phi_{t_0,t}\circ \gamma)]\\
    &\leq e^{-p\lambda(t-t_0)} \ell_{t_0}^p(\gamma).
\end{align*}
Taking infimum over all $\gamma\in\Gamma(\xi_1,\xi_2)$ then gives
\begin{align*}
    \bbE[\sfd_t^p(\phi_{t_0,t}(\xi_1),\phi_{t_0,t}(\xi_2))]
    &\leq e^{-p\lambda(t-t_0)} d_{t_0}^p(\xi_1,\xi_2).
\end{align*}
This completes the proof since $\phi_{t_0,t}(\xi_1)$ and $\phi_{t_0,t}(\xi_2)$ are the solutions of \eqref{eqn:sde} starting from $\xi_1$ and $\xi_2$ at time $t_0$.

\subsubsection*{Necessity}

We now prove Theorem~\ref{thm:Lp_contraction}-(b). Fix $(\xi,\zeta)\in T\bbR^n\setminus 0$.
Define a line segment $\gamma:[0,1]\rightarrow\bbR^n$ by $\gamma(r) \coloneqq \xi + r \zeta$, and define $x_r(t)$ and $v_r(t)$ as in \eqref{eqn:flow_of_curve} for each $r\in[0,1]$.
Since $(x_r(t))_{t\geq t_0}$ is a solution of \eqref{eqn:sde} for each $r\in[0,1]$, all driven by the same noise process, we have
\begin{equation*}
    \bbE[\sfd_t^p(x_0(t),x_r(t))] \leq e^{-p\lambda(t-t_0)} \sfd_{t_0}^p (x_0(t_0),x_r(t_0))
\end{equation*}
for each $t\geq t_0$ and each $r\in[0,1]$.

Recall that the distance between two nearby points can be approximated as \eqref{eqn:distance_approximation} under Hypothesis~\ref{hyp:convexity}. (See Lemma~\ref{lem:linear_approximation} in Appendix~\ref{appendix:technical_lemmas} for a precise statement.)
We thus have
\begin{equation*}
    \lim_{r\downarrow 0} \frac{\sfd_t(x_0(t),x_r(t))}{r} = F_t(x_0(t),v_0(t))
\end{equation*}
for each $t\geq t_0$.
Hence, invoking Fatou's lemma, we obtain
\begin{align*}
    \bbE\left[F_t^p(x_0(t),v_0(t))\right]
    &= \bbE\left[\lim_{r\downarrow 0} \frac{\sfd_t^p (x_0(t),x_r(t))}{r^p}\right]\\
    &\leq \liminf_{r\downarrow 0} \bbE\left[\frac{\sfd_t^p (x_0(t),x_r(t))}{r^p}\right]\\
    &\leq \liminf_{r\downarrow 0} e^{-p\lambda(t-t_0)}\frac{\sfd_{t_0}^p(x_0(t_0),x_r(t_0))}{r^p}\\
    &= e^{-p\lambda(t-t_0)}F_{t_0}^p(\xi,\zeta)
\end{align*}
for each $t\geq t_0$. In particular, this implies
\begin{equation}
\label{eqn:infinitesimal_distance_contraction_necessity}
    \limsup_{t\downarrow t_0} \frac{\bbE[F_t^p(x_0(t),v_0(t))] - F_{t_0}^p(\xi,\zeta)}{t-t_0} \leq -p\lambda F_{t_0}^p(\xi,\zeta).
\end{equation}

Now we introduce a stopping time
\begin{equation}
\label{eqn:stopping_time_necessity}
    \tau \coloneqq \inf\{t\geq t_0: \|x_0(t)-\xi\| \vee \|v_0(t)-\zeta\| \geq \|\zeta\|/2\}.
\end{equation}
Then, by It{\^o}'s formula, we have
\begin{equation}
\label{eqn:stopped_ito_necessity_proof}
    \bbE[F_{t\wedge\tau}^p(x_0,v_0)] = F_{t_0}^p(\xi,\zeta) + \int_{t_0}^{t} \bbE[1_{\{\tau>s\}}\scrL F_s^p(x_0,v_0)] ds,
\end{equation}
where $F_s(x_0,v_0)$ is shorthand for $F_s(x_0(s),v_0(s))$ and $1_{\{\tau>s\}}$ denotes the indicator for the event $\{\tau>s\}$, i.e., $1_{\{\tau>s\}}=1$ if $\tau>s$ and $1_{\{\tau>s\}}=0$ otherwise.
We claim
\begin{equation}
\label{eqn:stopped_generator}
    \lim_{t\downarrow t_0} \frac{\bbE[F_{t\wedge \tau}^p(x_0,v_0)]-F_{t_0}^p(\xi,\zeta)}{t-t_0} = \scrL F_{t_0}^p(\xi,\zeta).
\end{equation}
Indeed, by \eqref{eqn:stopped_ito_necessity_proof}, there exist $s_1,s_2\in[t_0,t]$ such that
\begin{align*}
    \frac{\bbE[F_{t\wedge \tau}^p(x_0,v_0)]-F_{t_0}^p(\xi,\zeta)}{t-t_0} &\leq \bbE[1_{\{\tau>s_1\}}\scrL F_{s_1}^p(x_0,v_0)],\\
    \frac{\bbE[F_{t\wedge \tau}^p(x_0,v_0)]-F_{t_0}^p(\xi,\zeta)}{t-t_0} &\geq \bbE[1_{\{\tau>s_2\}}\scrL F_{s_2}^p(x_0,v_0)].
\end{align*}
Under Hypothesis~\ref{hyp:regularity}, $1_{\{\tau>s_1\}} \scrL F_{s_1}^p(x_0,v_0) \rightarrow \scrL F_{t_0}^p(\xi,\zeta)$ as $t\downarrow t_0$ almost surely (see Lemma~\ref{lem:right_continuity} in Appendix~\ref{appendix:technical_lemmas}). Since $1_{\{\tau>s_1\}} \scrL F_{s_1}^p(x_0,v_0)$ is bounded, we further obtain $\lim_{t\downarrow t_0} \bbE[1_{\{\tau>s_1\}} \scrL F_{s_1}^p(x_0,v_0)] = \scrL F_{t_0}^p (\xi,\zeta)$
by the dominated convergence theorem.
Similarly, we also have $\lim_{t\downarrow t_0} \bbE[1_{\{\tau>s_2\}} \scrL F_{s_2}^p(x_0,v_0)] = \scrL F_{t_0}^p (\xi,\zeta)$.
Combining these gives \eqref{eqn:stopped_generator}.

Now it remains to show that
\begin{equation}
\label{eqn:stopped_generator_bound}
    \begin{aligned}
        &\lim_{t\downarrow t_0} \frac{\bbE[F_{t\wedge \tau}^p(x_0,v_0)]-F_{t_0}^p(\xi,\zeta)}{t-t_0}\\
        &\qquad \leq \limsup_{t\downarrow t_0} \frac{\bbE[F_{t}^p(x_0,v_0)]-F_{t_0}^p(\xi,\zeta)}{t-t_0}.
    \end{aligned}
\end{equation}
Since $F_{t\wedge \tau}^p(x_0,v_0) = 1_{\{\tau>t\}} F_t^p(x_0,v_0) + 1_{\{ \tau\leq t\}} F_{\tau}^p(x_0,v_0) \leq F_t^p(x_0,v_0) + C 1_{\{ \tau \leq t\}}$ for some constant $C>0$,
we have
\begin{equation*}
    \bbE[F_{t\wedge \tau}^p(x_0,v_0)]\leq \bbE[F_t^p(x_0,v_0)] + C \bbP(\tau\leq t).
\end{equation*}
Consequently, we obtain
\begin{align*}
    &\lim_{t\downarrow t_0} \frac{\bbE[F_{t\wedge \tau}^p(x_0,v_0)]-F_{t_0}^p(\xi,\zeta)}{t-t_0}\\
    &\leq \limsup_{t\downarrow t_0} \left[\frac{\bbE[F_{t}^p(x_0,v_0)]-F_{t_0}^p(\xi,\zeta)}{t-t_0} + C \frac{\bbP(\tau\leq t)}{t-t_0}\right].
\end{align*}
As the last term on the right-hand side vanishes by Lemma~\ref{lem:probability_vanish} in Appendix~\ref{appendix:technical_lemmas}, we obtain \eqref{eqn:stopped_generator_bound}.

Combining \eqref{eqn:infinitesimal_distance_contraction_necessity}, \eqref{eqn:stopped_generator}, and \eqref{eqn:stopped_generator_bound} now gives $\scrL F_{t_0}^p (\xi,\zeta) \leq -p\lambda F_{t_0}^p(\xi,\zeta)$.
Since $(t_0,\xi,\zeta)\in\bbR\times(T\bbR^n\setminus0)$ was arbitrary, the proof is complete.

\section{Almost Sure Contraction}
\label{sec:almost_sure_contraction}

\subsection{The Notion of Almost Sure Contraction}

We now turn to another mode of contraction. Extending the definition in \cite{monm23stoch} to general Finsler distances, we define almost sure contraction as follows.

\begin{definition}
\label{def:as_contraction}
    Stochastic dynamical system \eqref{eqn:sde} is said to be \emph{almost surely contracting (AS-contracting)} in distance $\sfd_t$ with rate $\lambda>0$ if
    \begin{equation}
    \label{eqn:as_contraction}
        \sfd_t(x_1(t),x_2(t)) \leq e^{-\lambda(t-t_0)} \sfd_{t_0}(\xi_1,\xi_2)\quad \forall t \geq t_0 \quad \text{a.s.}
    \end{equation}
    for every $t_0\in\bbR$ and $x_i(t_0)=\xi_i\in\bbR^n$, $i=1,2$.
\end{definition}

One can observe that the OU process \eqref{eqn:ou_process} is AS-contracting in the Euclidean distance if and only if $\alpha<0$, whereas the geometric Brownian motion \eqref{eqn:geometric_brownian_motion} cannot be AS-contracting in the Euclidean distance unless $\beta=0$ and $\alpha<0$.

In general, almost sure convergence and $L^p$-convergence of a family of random variables are incomparable; neither of them implies the other.
In our case, however, AS-contraction always implies $L^p$-contraction.
This is because Definition~\ref{def:as_contraction} provides a uniform control over the transient behavior.

\begin{proposition}
\label{prop:AS_hierarchy}
    If system~\eqref{eqn:sde} is AS-contracting, then it is $L^p$-contracting for every $p\in[1,\infty)$ in the same distance with the same rate.
\end{proposition}

\begin{proof}
    It follows immediately from Definitions~\ref{def:Lp_contraction} and~\ref{def:as_contraction}.
\end{proof}

\begin{remark}
    The notion of AS-contraction can be viewed as the limiting case of $L^p$-contraction. Indeed, if we replace the left-hand side of \eqref{eqn:Lp_contraction} with $\esssup \sfd_t(x_1(t),x_2(t))$, which is the $L^\infty$-norm of $\sfd_t(x_1(t),x_2(t))$, then the resulting condition coincides with \eqref{eqn:as_contraction} under the technical assumption that $\sfd_t(x,x')$ is jointly continuous in $(t,x,x')$ with respect to the Euclidean topology.
    This assumption holds, for example, when $\sfd_t$ is a time-independent Finsler distance.
\end{remark}

\subsection{Differential Characterization}
\label{sec:as_differential_characterization}

We now derive a differential characterization of AS-contraction.
Again, we consider two nearby trajectories $x(t)$ and $x(t)+v(t)$, as in Section~\ref{sec:Lp_differential_characterization}.
Then, with the approximation \eqref{eqn:distance_approximation} at hand, AS-contraction requires
\begin{equation}
\label{eqn:as_infinitesimal_contraction}
    F_t^p(x(t),v(t))\leq e^{-p\lambda(t-t_0)} F_{t_0}^p (\xi,\zeta) \quad \forall t\geq t_0 \quad \text{a.s.}
\end{equation}
for an arbitrary $p>0$.

Recalling that $x$ and $v$ are governed by the variational dynamics \eqref{eqn:lifted_sde}, an application of It{\^o}'s formula gives
\begin{equation}
\label{eqn:as_ito_formula}
\begin{aligned}
    F_t^p(x(t),v(t)) = F_{t_0}^p(\xi,\zeta) + \int_{t_0}^t \scrL F_s^p(x(s),v(s))ds&\\
    + \sum_{j=1}^m\int_{t_0}^t \left(\frac{\partial F_s^p}{\partial x}\sigma_j+ \frac{\partial F_s^p}{\partial v} \frac{\partial \sigma_j}{\partial x} v\right) dW_s^j&.
\end{aligned}
\end{equation}
In order for \eqref{eqn:as_infinitesimal_contraction} to hold, the stochastic integral in \eqref{eqn:as_ito_formula} should vanish; otherwise, it can take a sufficiently large positive value with nonzero probability, preventing $F_t^p(x(t),v(t))$ from satisfying \eqref{eqn:as_infinitesimal_contraction}.
This motivates the condition
\begin{equation}
\label{eqn:killing}
    \frac{\partial F_t^p}{\partial x}\sigma_j + \frac{\partial F_t^p}{\partial v}\frac{\partial \sigma_j}{\partial x}v=0,\quad j=1,\dots,m.
\end{equation}
Suppose now that \eqref{eqn:killing} holds.
Then, it is easy to see that the condition for $L^p$-contraction \eqref{eqn:finsler_condition} implies \eqref{eqn:as_infinitesimal_contraction}.

The discussion above identifies \eqref{eqn:finsler_condition} together with \eqref{eqn:killing}, for any $p>0$, as a candidate sufficient condition for AS-contraction.
Again, we rigorously prove its sufficiency and necessity under suitable hypotheses.

\begin{theorem}
\label{thm:AS_contraction}
    Let $F_t(\cdot,\cdot)$ be a Finsler structure and $\sfd_t(\cdot,\cdot)$ denote the induced Finsler distance.
    Suppose that Hypotheses~\ref{hyp:regularity} and~\ref{hyp:smoothness} hold.
    Let $\lambda>0$.
    \begin{enumerate}[label=(\alph*),nosep]
        \item If \eqref{eqn:finsler_condition} and \eqref{eqn:killing} hold for all $(t,x,v)\in\bbR\times (T\bbR^n\setminus0)$ for some $p> 0$, then system~\eqref{eqn:sde} is AS-contracting in $\sfd_t$ with rate $\lambda$.
        \item If system \eqref{eqn:sde} is AS-contracting in $\sfd_t$ with rate $\lambda$, and $F_t$ satisfies Hypothesis~\ref{hyp:convexity}, then \eqref{eqn:finsler_condition} and~\eqref{eqn:killing} hold for all $(t,x,v)\in\bbR\times (T\bbR^n\setminus0)$ and all $p>0$.
    \end{enumerate}
\end{theorem}

\begin{proof}
    See Section~\ref{sec:proof_as}.
\end{proof}

Several remarks on conditions~\eqref{eqn:finsler_condition} and \eqref{eqn:killing} are in order.

\begin{enumerate}[label=(\roman*),resume,nosep]
    \item 
    Viewing the noise as an external input acting through $\sigma_j$, condition \eqref{eqn:killing} states that this input has no effect on the Finsler distance.
    Indeed, the left-hand side of \eqref{eqn:killing} represents the rate of change of $F_t^p(x,v)$ when $x$ and $x+v$ evolve along the vector field $\sigma_j$.
    The motivation behind condition~\eqref{eqn:killing} can thus be understood as follows: If \eqref{eqn:killing} does not hold, a noise realization of sufficiently large magnitude and suitable sign, which occurs with nonzero probability, could increase the Finsler distance.
    
    \item
    \label{item:as_condition_equivalence}
    For any $q>0$, the conditions obtained by replacing $p$ with $q$ in \eqref{eqn:finsler_condition} and \eqref{eqn:killing} are equivalent to the original conditions.
    Indeed, by the chain rule, we have
    \begin{equation*}
        \frac{\partial F_t^p}{\partial x}\sigma_j + \frac{\partial F_t^p}{\partial v}\frac{\partial \sigma_j}{\partial x}v=0 \:\: \Leftrightarrow \:\: \frac{\partial F_t^q}{\partial x}\sigma_j + \frac{\partial F_t^q}{\partial v}\frac{\partial \sigma_j}{\partial x}v=0.
    \end{equation*}
    Furthermore, once this condition is satisfied, the quantity $\tilde \Delta_{j,p,q}$ defined in item~\ref{item:another_representation} vanishes.
    Thus, in view of \eqref{eqn:another_finsler_condition}, condition \eqref{eqn:finsler_condition} with exponent $p$ becomes equivalent to that with exponent $q$.
    This allows one to choose a favorable exponent $q$ for which the computation of $F_t^q$ is simpler.

    \item
    Item~\ref{item:as_condition_equivalence} also shows that the differential conditions for AS-contraction imply the differential condition for $L^p$-contraction for every $p\in[1,\infty)$.
    This is consistent with Proposition~\ref{prop:AS_hierarchy}.

    \item Items~\ref{item:comparison_with_deterministic_contraction}, \ref{item:noise_induced_contraction}, and \ref{item:unit_tangent_bundle}--\ref{item:locally_bounded} remain valid for AS-contraction. For item~\ref{item:comparison_with_deterministic_contraction}, observe that condition~\eqref{eqn:killing} becomes trivial in the noise-free case.
    Item~\ref{item:noise_induced_contraction} remains valid since $\Delta_{j,p}$ can still be negative while condition~\eqref{eqn:killing} is satisfied.
    Regarding item~\ref{item:unit_tangent_bundle}, it can be seen that all terms on the left-hand side of \eqref{eqn:killing} are positively homogeneous of degree $p$ in $v$.
\end{enumerate}

\subsection{Proof of Theorem~\ref{thm:AS_contraction}}
\label{sec:proof_as}
\subsubsection*{Sufficiency}

The proof of Theorem~\ref{thm:AS_contraction}-(a) proceeds in the same three steps as in the proof of Theorem~\ref{thm:Lp_contraction}-(a).

\textit{Step 1:}
As discussed in Section~\ref{sec:as_differential_characterization}, we have infinitesimal contraction property~\eqref{eqn:as_infinitesimal_contraction} under conditions~\eqref{eqn:finsler_condition} and \eqref{eqn:killing}.

\textit{Step 2:} 
Let $\gamma:[0,1]\rightarrow\bbR^n$ be a regular piecewise $C^1$ curve.
Define $x_r(t)$ and $v_r(t)$ for all $r\in [0,1]$ as in \eqref{eqn:flow_of_curve}.
Then, by Step 1, we have
\begin{equation}
\label{eqn:as_infinitesimal_displacement_contraction}
    F_t(x_r(t),v_r(t)) \leq e^{-\lambda(t-t_0)} F_{t_0}(\gamma(r),\dot\gamma(r)) \quad \forall t \geq t_0
\end{equation}
for all $r\in \bbQ \cap [0,1]$ almost surely.
In fact, \eqref{eqn:as_infinitesimal_displacement_contraction} holds for all $r\in[0,1]$ almost surely because $F_t^p(x_r(t),v_r(t))$ is piecewise continuous in $r$.
Note that both $x_r(t)$ and $v_r(t)$ are piecewise continuous in $r$ by Lemma~\ref{lem:stochastic_flow} and that $F_t(x,v)$ is continuous in $(x,v)$ under Hypothesis~\ref{hyp:smoothness}.
We can now conclude that
\begin{align*}
    \ell_t(\phi_{t_0,t} \circ \gamma) &= \int_0^1 F_t(x_r(t),v_r(t)) dr\\
    &\leq e^{-\lambda(t-t_0)} \int_0^1 F_{t_0}(\gamma(r), \dot\gamma(r))dr\\
    &=e^{-\lambda(t-t_0)} \ell_{t_0}(\gamma)
\end{align*}
holds for all $t\geq t_0$ almost surely.

\textit{Step 3:} 
Fix $\xi_1,\xi_2\in\bbR^n$, and choose a sequence of regular piecewise $C^1$ curves $(\gamma_n)_{n\in\bbN}\subseteq \Gamma(\xi_1,\xi_2)$ such that
\begin{equation*}
    \ell_{t_0}(\gamma_n) \leq \sfd_{t_0}(\xi_1,\xi_2) + \frac{1}{n}.
\end{equation*}
Applying the result of Step 2, for each $n\in\bbN$, we obtain
\begin{align*}
    &\sfd_{t}(\phi_{t_0,t}( \xi_1),\phi_{t_0,t}(\xi_2)) \leq \ell_t(\phi_{t_0,t}\circ \gamma_n)\\
    &\qquad \leq e^{-\lambda(t-t_0)} \ell_{t_0}(\gamma_n) \leq e^{-\lambda(t-t_0)} \left(\sfd_{t_0}(\xi_1,\xi_2) + \frac{1}{n} \right)
\end{align*}
for all $t\geq t_0$ almost surely.
Since $\bbN$ is countable, we have
\begin{equation*}
    \sfd_t(\phi_{t_0,t}(\xi_1),\phi_{t_0,t}(\xi_2)) \leq e^{-\lambda(t-t_0)} \sfd_{t_0}(\xi_1,\xi_2) \quad \forall t \geq t_0
\end{equation*}
almost surely.

\subsubsection*{Necessity}

Suppose that \eqref{eqn:sde} is AS-contracting.
Then, by Proposition~\ref{prop:AS_hierarchy}, it is $L^p$-contracting for every $p\in[1,\infty)$.
Consequently, by Theorem~\ref{thm:Lp_contraction}-(b), \eqref{eqn:finsler_condition} holds for all $(t,x,v)\in\bbR\times (T\bbR^n\setminus 0)$ and all $p\in[1,\infty)$.

Suppose to the contrary that \eqref{eqn:killing} does not hold for some $q>0$.
Then, there exists $(t,x,v)\in\bbR\times(T\bbR^n\setminus 0)$ such that
\begin{equation*}
    \left(\frac{\partial F_t^q}{\partial x}\sigma_j + \frac{\partial F_t^q}{\partial v}\frac{\partial \sigma_j}{\partial x}v\right)^2 > 0.
\end{equation*}
In view of \eqref{eqn:another_finsler_condition}, this implies that \eqref{eqn:finsler_condition} cannot hold for large values of $p$.
This completes the proof.

\section{Special Cases}
\label{sec:special_cases}

The established differential conditions can be specialized to Riemannian and norm-induced distances.
Notably, the resulting conditions are both necessary and sufficient for contraction since the associated Finsler structures are convex.

\subsection{Contraction in a Riemannian Distance}

Let $M_t(x)\in\bbR^{n\times n}$ be a Riemannian metric that induces a Finsler structure $F_t(x,v) = \sqrt{v^\top M_t(x) v}$ as in Example~\ref{ex:riemannian}.
Denoting the $ij$-th component of $M_t(x)$ by $M_t^{ij}(x)$, we define the directional derivatives of $M_t$ as follows.
For $\eta\in\bbR^n$, $D_xM_t(x)[\eta]\in\bbR^{n\times n}$ denotes the component-wise directional derivative of $M_t(x)$ with respect to $x$ in  direction $\eta$, i.e., its $ij$-th component is given by $(\partial_x M_t^{ij}(x)) \eta$.
Similarly, $D_x^2 M_t(x)[\eta,\eta]\in\bbR^{n\times n}$ denotes the component-wise second directional derivative of $M_t(x)$ with respect to $x$ in direction $\eta$, i.e., its $ij$-th component is given by $\eta^\top (\partial_x^2 M_t^{ij}(x))\eta$.

We now provide a specialized condition for Riemannian cases.
Since the Finsler structure induced by a Riemannian metric satisfies Hypothesis~\ref{hyp:convexity}, Theorems~\ref{thm:Lp_contraction} and \ref{thm:AS_contraction} yield the following necessary and sufficient condition for $L^2$- and AS-contraction in a Riemannian distance.

\begin{corollary}
\label{cor:riemannian}
    Suppose that Hypothesis~\ref{hyp:regularity} holds.
    Let $M_t(\cdot)$ be a Riemannian metric and $\sfd_t(\cdot,\cdot)$ be the induced Riemannian distance.
    Suppose that the partial derivatives
    \begin{equation*}
        \frac{\partial M_t^{ij}}{\partial t}, \frac{\partial M_t^{ij}}{\partial x},\frac{\partial^2 M_t^{ij}}{\partial x^2}, \quad i,j=1,\dots,n,
    \end{equation*}
    exist as continuous functions in $(t,x)$ on $\bbR\times\bbR^n$.
    \begin{enumerate}[label=(\alph*),nosep]
        \item System \eqref{eqn:sde} is $L^2$-contracting in the Riemannian distance $\sfd_t$ with rate $\lambda>0$ if and only if
        \begin{equation}
        \label{eqn:riemannian_condition}
            \frac{\partial M_t}{\partial t} + D_x M_t [f] + \frac{\partial f}{\partial x}^{\!\top} \!M_t + M_t \frac{\partial f}{\partial x} + \sum_{j=1}^m \hat\Delta_j \preceq -2\lambda M_t
        \end{equation}
        for all $(t,x)\in\bbR\times \bbR^n$, where $\hat\Delta_j(t,x)\in\bbR^{n\times n}$ denotes
        \begin{equation*}
        \begin{aligned}
            \hat\Delta_j(t,x) \coloneqq \frac{1}{2} D_x^2 M_t [\sigma_j,\sigma_j] + D_x M_t [\sigma_j]\frac{\partial \sigma_j}{\partial x}&\\
            \quad+ \frac{\partial \sigma_j}{\partial x}^{\!\top}\! D_x M_t[\sigma_j]+ \frac{\partial \sigma_j}{\partial x}^{\!\top}\! M_t \frac{\partial \sigma_j}{\partial x}&
        \end{aligned}
        \end{equation*}
        for each $j=1,\dots,m$.
        \item System~\eqref{eqn:sde} is AS-contracting in the Riemannian distance~$\sfd_t$ with rate $\lambda>0$ if and only if both \eqref{eqn:riemannian_condition} and
        \begin{equation}
        \label{eqn:killing_Riemannian}
            D_xM_t[\sigma_j] + \frac{\partial \sigma_j}{\partial x}^{\!\top} \!M_t + M_t \frac{\partial \sigma_j}{\partial x} = 0,\:\: j=1,\dots,m,
        \end{equation}
        hold for all $(t,x)\in\bbR\times\bbR^n$.
    \end{enumerate}
\end{corollary}

Some remarks are in order.

\begin{itemize} [nosep]
    \item In the noise-free case, condition \eqref{eqn:riemannian_condition} reduces exactly to the deterministic contraction condition in \cite{lohmiller98aut}. Condition \eqref{eqn:killing_Riemannian} becomes trivial.

    \item
    Note that the noise-related terms $\hat \Delta_{j}$ can be sign indefinite or even negative definite.
    Consequently, system~\eqref{eqn:sde} can be $L^2$- or AS-contracting in a Riemannian distance even if the noise-free system $\dot x = f(t,x)$ is not contracting.

    \item For $p\neq 2$, it is generally not possible to formulate a matrix-valued condition that characterizes $L^p$-contraction in a Riemannian distance. Still, one can use condition~\eqref{eqn:another_finsler_condition} with $q=2$ to obtain a necessary and sufficient condition for $L^p$-contraction in a Riemannian distance.    

    \item
    A generalization of condition~\eqref{eqn:riemannian_condition} to Riemannian manifolds appears in the concurrent work \cite{dongjun} on Wasserstein contraction.
    Specifically, the author established that the generalized condition is sufficient for $L^2$-contraction on geodesically complete Riemannian manifolds\footnote{To clarify, the Riemannian structures on $\bbR^n$ considered in this paper are not assumed to be geodesically complete.}
    and then connected it to 2-Wasserstein contraction.
    
\end{itemize}

\subsection{Contraction in a Norm}

We now specialize the preceding results to norm-induced distances.
Again, the convexity of the associated Finsler structure leads to necessary and sufficient conditions.

\begin{corollary}
\label{cor:norm}
    Suppose that Hypothesis~\ref{hyp:regularity} holds.
    Let $|\cdot|$ be a norm which is $C^2$ on $\bbR^n\setminus\{0\}$.
    \begin{enumerate}[label=(\alph*)]
        \item Let $p\in[1,\infty)$. System \eqref{eqn:sde} is $L^p$-contracting in the norm $|\cdot|$ with rate $\lambda>0$ if and only if
        \begin{equation}
        \label{eqn:norm_condition}
        \begin{aligned}
            \frac{\partial |v|^p}{\partial v} \frac{\partial f}{\partial x} v \!+ \!\frac{1}{2}\sum_{j=1}^m \!\left(\frac{\partial \sigma_j}{\partial x} v\!\right)^{\!\!\!\top} \!\!\left[\frac{\partial^2 |v|^p}{\partial v^2}\right]\!\!\left(\frac{\partial \sigma_j}{\partial x}v\!\right) \!\leq\! -p\lambda |v|^p
        \end{aligned}
        \end{equation}
        for all $(t,x,v)\in\bbR\times (T\bbR^n\setminus0)$.
        \item Let $p > 0$. System \eqref{eqn:sde} is AS-contracting in the norm $|\cdot|$ with rate $\lambda>0$ if and only if \eqref{eqn:norm_condition} and
        \begin{equation}
        \label{eqn:killing_norm}
            \frac{\partial |v|}{\partial v} \frac{\partial \sigma_j}{\partial x}v =0,\quad j=1,\dots,m,
        \end{equation}
        hold for all $(t,x,v)\in\bbR\times (T\bbR^n\setminus0)$.
    \end{enumerate}
\end{corollary}

Some remarks are in order.
\begin{itemize} [nosep]
    \item In the noise-free case, condition \eqref{eqn:killing_norm} becomes trivial, and condition \eqref{eqn:norm_condition} becomes
    \begin{equation}
    \label{eqn:norm_condition_deterministic}
        \frac{\partial |v|}{\partial v} \frac{\partial f}{\partial x}v \leq -\lambda |v|.
    \end{equation}
    When $|\cdot|$ is differentiable on $\bbR^n\setminus\{0\}$, \eqref{eqn:norm_condition_deterministic} is equivalent to the deterministic contraction condition $\mu(\frac{\partial f}{\partial x}) \leq -\lambda$ discussed in \cite{sontag2010,aminzare2014cdc}, where $\mu$ denotes the logarithmic norm (or the matrix measure) associated with $|\cdot|$.
    For further details, see \cite[Lemmas~4 and~5]{davydov2022tac}. See also \cite[Sec.~V-A]{FS14}.
    
    \item When the effect of noise is independent of the state, i.e., $\sigma_j(t,x)=\sigma_j(t)$, again, condition~\eqref{eqn:killing_norm} becomes trivial, and condition~\eqref{eqn:norm_condition} reduces to \eqref{eqn:norm_condition_deterministic}.
    Thus, we can conclude that deterministic contraction, $L^p$-contraction for every $p\in[1,\infty)$, and AS-contraction are \emph{all equivalent} for state-independent noise and norm-induced distances.
    This generalizes the result in \cite{monm23stoch} to general $C^2$ norms and time-varying coefficients.
    
    \item The noise-related term in \eqref{eqn:norm_condition} is always nonnegative for $p\geq1$.
    Consequently, in order for a system to be $L^p$- or AS-contracting in a norm, the noise-free system $\dot x = f(t,x)$ must also be contracting in the same norm.\footnote{The result in \cite[Cor.~2]{amin22lcss}, according to which multiplicative noise enhances $L^p$-contraction in a norm, conflicts with ours and does not appear to be valid as stated.
    The geometric Brownian motion~\eqref{eqn:geometric_brownian_motion} with $\alpha=0$ and $\beta=1$ provides a counterexample: It satisfies the hypotheses of \cite[Cor.~2]{amin22lcss}, yet by Example~\ref{ex:geometric_brownian_motion} it is not $L^p$-contracting for any $p\geq 1$.}

    \item Previous studies have introduced stochastic logarithmic norms \cite{ahmad2010} and stochastic logarithmic Lipschitz constants \cite{amin22lcss} for verifying $L^p$-contraction in a norm.
    However, these quantities are difficult to compute directly, and their use often relies on estimates.
    Condition \eqref{eqn:norm_condition} provides an alternative, readily computable criterion for verifying stochastic contraction in a norm.
\end{itemize}

\section{Further Discussions}
\label{sec:further_discussions}

\subsection{Connections to Related Notions}

Contraction properties of stochastic systems have been defined in various ways in the literature, sometimes under the same terminology.
Here, we investigate connections between our definitions and other definitions in the literature.

In \cite{liu21tac,ngoc24ijc}, the following property, which we shall call almost sure incremental exponential stability,\footnote{This terminology is motivated by the definition of almost sure exponential stability in \cite[Sec.~4.3]{Mao08}, extended here to its incremental counterpart.} is termed almost sure contraction.

\begin{definition}
    Stochastic dynamical system \eqref{eqn:sde} is said to be \emph{almost surely incrementally exponentially stable (AS-IES)} in distance $\sfd_t$ with rate $\lambda>0$ if
    \begin{equation*}
    \label{eqn:as_ies}
        \limsup_{t\rightarrow\infty} \frac{\ln \sfd_t(x_1(t),x_2(t))}{t} \leq -\lambda \quad \text{a.s.}
    \end{equation*}
    for every $t_0\in\bbR$ and $x_i(t_0)=\xi_i\in\bbR^n$, $i=1,2$.
\end{definition}

This is a weaker property that is implied by AS-contraction and even by $L^p$-contraction for every $p\in[1,\infty)$.

\begin{proposition}
\label{prop:as_ies_hierarchy}
    Let $\sfd_t(x,x')$ be a distance function that is jointly continuous in $(t,x,x')$ with respect to the Euclidean topology.
    If system~\eqref{eqn:sde} is either AS-contracting in $\sfd_t$ or $L^p$-contracting in $\sfd_t$ for some $p\in[1,\infty)$, then it is AS-IES in $\sfd_t$ with the same rate.
\end{proposition}

\begin{proof}
    Since AS-contraction implies $L^p$-contraction by Proposition~\ref{prop:AS_hierarchy}, we may assume that \eqref{eqn:sde} is $L^p$-contracting. Then,
    for any $t\geq s \geq t_0$, we have
    \begin{equation*}
        \bbE [e^{p\lambda t} \sfd_{t}^p(x_1(t),x_2(t))|\calF_{s}] \leq e^{p\lambda s} \sfd_{s}^p(x_1(s),x_2(s))\quad \text{a.s.}
    \end{equation*}
    Thus, $e^{p\lambda t} \sfd_t^p(x_1(t),x_2(t))$ is a nonnegative supermartingale with continuous sample paths. Consequently, by the martingale convergence theorem \cite[Thm.~A.6]{schilling}, $\lim_{t\rightarrow \infty} e^{p \lambda t} \sfd_t^p(x_1(t),x_2(t))$ exists for $\bbP$-almost every $\omega\in\Omega$.
    Thus, we can find $C:\Omega\rightarrow\bbR$ such that
    \begin{equation}
    \label{eqn:as_contraction_exponential_bound}
        \sfd_t^p (x_1(t),x_2(t)) \leq C(\omega) e^{-p \lambda t} \quad \forall t\geq t_0
    \end{equation}
    for $\bbP$-almost every $\omega\in\Omega$.
    Therefore, we can conclude that
    \begin{equation*}
        \limsup_{t\rightarrow\infty}\frac{\ln \sfd_t^p(x_1(t),x_2(t))}{t} \leq -p\lambda + \limsup_{t\rightarrow\infty}\frac{\ln C(\omega)}{t} = - p\lambda
    \end{equation*}
    almost surely. This completes the proof.
\end{proof}

It is worth noting that AS-IES does not imply $L^p$-contraction.
Indeed, following \cite[Example~3.5.5]{Mao08}, one can show that the geometric Brownian motion~\eqref{eqn:geometric_brownian_motion} is AS-IES in the Euclidean distance if and only if $\alpha<\beta^2/2$.
This condition is strictly weaker than that for $L^p$-contraction (see Example~\ref{ex:geometric_brownian_motion}).
This is because AS-IES provides no control over transient behavior; $C(\omega)$ in \eqref{eqn:as_contraction_exponential_bound} may depend on the sample paths and need not be integrable, i.e., we may have $\bbE[C(\omega)]=\infty$.

\begin{remark}
    The notion of $L^p$-contraction can be extended to $p\in(0,1)$ by treating $\|\cdot\|_{L^p}$ as a quasinorm, and Proposition~\ref{prop:as_ies_hierarchy} remains valid in this range.
    Thus, AS-IES is strictly weaker than $L^p$-contraction even for $p\in(0,1)$.
\end{remark}

The following notions of \emph{uniform} incremental exponential stability have also been considered in the literature.

\begin{definition}
\label{def:uies}
    Stochastic dynamical system \eqref{eqn:sde} is said to be
    \begin{enumerate}[label=(\alph*),nosep]
        \item \emph{$L^p$-uniformly incrementally exponentially stable ($L^p$-UIES)} in distance $\sfd_t$ with rate $\lambda>0$ if there exists $C>0$ such that
        \begin{equation*}
            \|\sfd_t(x_1(t),x_2(t))\|_{L^p} \leq C e^{-\lambda(t-t_0)}\sfd_{t_0}(\xi_1,\xi_2) \quad \forall t\geq t_0
        \end{equation*}
        for every $t_0\in\bbR$ and $x_i(t_0)=\xi_i\in\bbR^n$, $i=1,2$,   
        
        \item \emph{almost surely uniformly incrementally exponentially stable (AS-UIES)} in distance $\sfd_t$ with rate $\lambda>0$ if there exists $C>0$ such that
        \begin{equation*}
            \sfd_t(x_1(t),x_2(t)) \leq C e^{-\lambda(t-t_0)}\sfd_{t_0}(\xi_1,\xi_2)\quad \forall t \geq t_0 \quad \text{a.s.}
        \end{equation*}
        for every $t_0\in\bbR$ and $x_i(t_0)=\xi_i\in\bbR^n$, $i=1,2$.
    \end{enumerate}
\end{definition}

To clarify, $L^p$-UIES is termed uniform incremental exponential stability in the $p$-th moment in \cite{kawano24tac,kawano25tac} for discrete-time systems and $p$-th moment contraction in \cite{ngoc24ijc}.
Regarding AS-UIES, note that it provides uniform control over transient behavior unlike AS-IES.

Clearly, $L^p$- and AS-contraction imply $L^p$- and AS-UIES, respectively.
In addition, each contraction property implies the corresponding UIES property for any equivalent distances.

\begin{proposition}
\label{prop:equivalent_distance}
    Let $\sfd_t$ and $\sfd_t'$ be equivalent distance functions, i.e., there exist constants $c_1,c_2>0$ such that
    \begin{equation}
    \label{eqn:equivalent_distances}
        c_1 d'_t(x,x') \leq d_t(x,x')\leq c_2 d'_t(x,x')
    \end{equation}
    for all $t\in\bbR$ and $x,x'\in\bbR^n$.
    If system~\eqref{eqn:sde} is $L^p$-contracting or AS-contracting in $\sfd_t$ with rate $\lambda>0$, then it is $L^p$-UIES or AS-UIES in $\sfd'_t$ with rate $\lambda$, respectively. The constant $C$ in Definition~\ref{def:uies} is given by $C=c_2/c_1$.
\end{proposition}

\begin{proof}
    The results follow directly from the definitions.
\end{proof}

If $\sfd_t$ and $\sfd_t'$ are the Finsler distances induced by Finsler structures $F_t$ and $F_t'$, respectively, \eqref{eqn:equivalent_distances} is implied by
\begin{equation*}
    c_1 F_t'(x,v) \leq F_t(x,v)\leq c_2 F_t'(x,v) \quad \forall(t,x,v)\in\bbR\times T\bbR^n.
\end{equation*}

In practice, estimates in the Euclidean distance are often of primary interest.
For this, one may need to verify that the Finsler structure certifying contraction of a system is uniform, that is, there exist constants $c_1,c_2>0$ such that
 \begin{equation*}
    c_1 \|v\|_2 \leq F_t(x,v)\leq c_2\|v\|_2 \quad \forall (t,x,v)\in\bbR\times T\bbR^n.
\end{equation*}
Indeed, if a Finsler structure $F_t$ is uniform, then the induced distance $\sfd_t$ is equivalent to the Euclidean distance, and consequently, if system~\eqref{eqn:sde} is contracting in the induced distance $\sfd_t$, then it is UIES in the Euclidean distance.

\begin{remark}
    In the concurrent work \cite{dongjun}, a condition closely related to \eqref{eqn:finsler_condition} emerges as a sufficient condition for $L^p$-UIES in the Euclidean distance, $p\in[1,2]$, in the course of establishing Wasserstein contraction.
    While the present paper uses the diffeomorphic property of the stochastic flow, \cite{dongjun} employs mean-square derivatives. 
    Both approaches yield sufficient conditions for $L^p$-UIES (and Wasserstein contraction) in the Euclidean distance under different technical requirements, neither of which subsumes the other.
\end{remark}

\subsection{The Role of Convexity}
\label{sec:the_role_of_convexity}

We now discuss why the convexity of a Finsler structure (Hypothesis~\ref{hyp:convexity}) is needed for the differential conditions to be necessary.
Motivated by \cite[Example~2.4.4]{BBI01}, we consider the following nonconvex Finsler structure.

\begin{figure}[t]
    \centering
    \includegraphics[width=0.8\columnwidth]{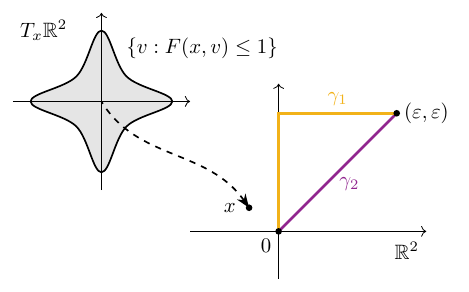}
    \caption{A nonconvex unit ball and curves connecting $(0,0)$ and $(\varepsilon,\varepsilon)$}
    \label{fig:role_of_convexity}
\end{figure}

\begin{example}
Let $F:T\bbR^2\rightarrow\bbR$ be a Finsler structure such that $F(x,(1,0))=F(x,(0,1))=1/2,\quad F(x,(1,1))=\sqrt{2}$
so that the Finsler unit ball $\{v:F(x,v)\leq 1\}\subseteq T_x\bbR^2$ has the star-shaped geometry shown in Fig.~\ref{fig:role_of_convexity}.
Now consider two regular piecewise $C^1$ curves $\gamma_1$ and $\gamma_2$ depicted in Fig.~\ref{fig:role_of_convexity} that connect the origin and $(\varepsilon,\varepsilon)$ for a small $\varepsilon>0$.
Then, we have $\ell(\gamma_1)=\varepsilon$ and $\ell(\gamma_2)=\sqrt{2}\varepsilon$.
In particular, we have $\sfd(0,(\varepsilon,\varepsilon))\leq \ell(\gamma_1)<\ell(\gamma_2)$,
where $\sfd(\cdot,\cdot)$ denotes the Finsler distance induced by $F$.
\end{example}

For this example, we can infer that the contraction property of the two nearby trajectories starting from $0$ and $(\varepsilon,\varepsilon)$ would depend on the contraction of $F(0,(0,\varepsilon))$ and $F(0,(\varepsilon,0))$, but not on that of $F(0,(\varepsilon,\varepsilon))$.
More generally, we could infer that when the Finsler structure is nonconvex, certain directions may not contribute to the induced distance, and consequently, contraction of infinitesimal displacements in every direction is not necessary to guarantee contraction in the induced distance.

When the Finsler structure is convex, in contrast, the distance between two nearby points $x$ and $x+v$ can always be approximated by $F(x,v)$.
Thus, we need contraction property of $F(x,v)$ for all directions of $v$, which in turn makes \eqref{eqn:finsler_condition} and \eqref{eqn:killing} necessary for all $(t,x,v)\in\bbR\times(T\bbR^n\setminus0)$.

\subsection{Incremental Input-to-State Stability}
\label{sec:iss}

A robustness property naturally expected of a contracting system is incremental input-to-state stability ($\delta$ISS) \cite{davydov2022tac,angeli2002tac,angeli2009tac,kawano2026arxiv}.
To study $\delta$ISS in the present setting, we consider two trajectories driven by common noise
\begin{subequations}
    \label{eqn:iss}
    \begin{align}
        dx_1 &= f(t,x_1,0)dt + \sigma(t,x_1)dW_t, \label{eqn:iss_no_input}\\
        dx_2 &= f(t,x_2,u)dt + \sigma(t,x_2)dW_t, \label{eqn:iss_with_input}
    \end{align}
\end{subequations}
where $x_1$ is uncontrolled and $x_2$ is controlled by an $(\calF_t)_{t\geq t_0}$-adapted input process $u(t)\in\bbR^{n_u}$.
We assume that both \eqref{eqn:iss_no_input} and \eqref{eqn:iss_with_input} admit a unique global strong solution for every initial time and initial condition.
We also impose the following hypothesis.

\begin{hypothesis}
\label{hyp:iss_regularity}
    There exist constants $L_x,L_u>0$ such that
    \begin{align*}
        (x-x')^\top (f(t,x,0)-f(t,x',0)) &\leq L_x\|x-x'\|_2^2\\
        \|\sigma(t,x) - \sigma(t,x')\|_F &\leq L_x \|x-x'\|_2\\
        \|f(t,x,u)-f(t,x,0)\|_2 &\leq L_u \|u\|_2
    \end{align*}
    for all $t\in\bbR$, $x,x'\in\bbR^n$, and $u\in\bbR^{n_u}$.
\end{hypothesis}

Regarding the input process, we introduce the following hypothesis, which is satisfied when $u(t)$ is a deterministic locally integrable signal, or when $u(t)$ is a solution of an SDE under standard regularity assumptions \cite[Sec.~2.4]{Mao08}.
Denote $\|u(t)\|_{L^p}\coloneqq\| \|u(t)\|_2\|_{L^p}$.

\begin{hypothesis}
\label{hyp:input}
    The input process $u(t)$ satisfies
    \begin{equation*}
        \int_{t_0}^t \|u(s)\|_{L^p} ds < \infty
    \end{equation*}
    for all $t_0\in\bbR$ and $t\geq t_0$.
\end{hypothesis}

Under these hypotheses, we establish the following $\delta$ISS result.

\begin{theorem}
\label{thm:iss}
    Let $p \geq 1$.
    Suppose that Hypotheses~\ref{hyp:iss_regularity} and \ref{hyp:input} hold.
    Suppose that $\sfd_t$ is a distance function satisfying
    \begin{equation}
    \label{eqn:iss_distance_upper_bound}
        \sfd_t(x,x')\leq C\|x-x'\|_2 \quad \forall t\in\bbR,\:\forall x,x'\in\bbR^n
    \end{equation}
    for some constant $C>0$.
    If system \eqref{eqn:iss_no_input} is $L^p$-contracting in $\sfd_t$ with rate $\lambda>0$, then we have
    \begin{equation}
    \label{eqn:iss_bound}
    \begin{aligned}
        &\|\sfd_t(x_1(t),x_2(t))\|_{L^p}\\
        &\leq e^{-\lambda(t-t_0)}\sfd_{t_0}(\xi_1,\xi_2)+ CL_u\int_{t_0}^t e^{-\lambda (t-s)} \|u(s)\|_{L^p}ds
    \end{aligned}
    \end{equation}
    for all $t\geq t_0$, for any $t_0\in\bbR$ and $x_i(t_0)=\xi_i\in\bbR^n$, $i=1,2$.
\end{theorem}

\begin{proof}

    The result is immediate when $t=t_0$.
    Fix $t>t_0$ and $K\in\bbN$.
    For $k=1,\dots,K$, define $t_k \coloneqq t_0 + {k}(t-t_0)/K$.
    For $k=0,1,\dots,K-1$, let $(y_k(s))_{s\geq t_k}$ be the solution of the SDE
    \begin{equation*}
        dy_k = f(s,y_k,0)ds + \sigma(s,y_k)dW_s
    \end{equation*}
    with initial condition $y_k(t_k)=x_2(t_k)$.
    Then, by the triangle inequality, we have
    \begin{equation*}
        \|\sfd_{t_{k+1}}(x_1,x_2)\|_{L^p} \leq \|\sfd_{t_{k+1}}(x_1,y_k)\|_{L^p} + \|\sfd_{t_{k+1}} (y_k,x_2)\|_{L^p}
    \end{equation*}
    for $k=0,\dots,K-1$. Here, we denote $\sfd_s(x,x')\coloneqq \sfd_s(x(s),x'(s))$ for notational conciseness.
    Since \eqref{eqn:iss_no_input} is $L^p$-contracting, the first term on the right-hand side can be estimated as
    \begin{equation*}
        \|\sfd_{t_{k+1}}(x_1,y_k)\|_{L^p} \leq e^{-\lambda (t_{k+1}-t_k)} \|\sfd_{t_k}(x_1,x_2)\|_{L^p}
    \end{equation*}
    by Proposition~\ref{prop:Lp_random_initial_condition}.
    By Lemma~\ref{lem:iss_lemma} in Appendix~\ref{appendix:technical_lemmas} and \eqref{eqn:iss_distance_upper_bound}, the second term on the right-hand side can be estimated as
     \begin{equation*}
        \|\sfd_{t_{k+1}}(y_k,x_2)\|_{L^p} \leq CL_u\int_{t_k}^{t_{k+1}} e^{M (t_{k+1} -s)}\|u(s)\|_{L^p}ds.
    \end{equation*}
    Combining the estimates for all $k$, we now obtain
    \begin{align*}
        &\|\sfd_{t}(x_1,x_2)\|_{L^p} \leq e^{-\lambda(t-t_0)}\sfd_{t_0}(\xi_1,\xi_2)\\
        & + CL_u \sum_{k=1}^{K} \int_{t_{k-1}}^{t_k} e^{-\lambda (t-t_k)+M(t_k-s)} \|u(s)\|_{L^p}ds.
    \end{align*}
    Write the last term as
    \begin{equation*}
       CL_u \int_{t_0}^t \psi_K(s)\|u(s)\|_{L^p}ds
    \end{equation*}
    for $\psi_K(s) \coloneqq \sum_{k=1}^{K} e^{-\lambda (t-t_k)+M(t_k-s)} 1_{[t_{k-1},t_k)}(s)$.
    Then, $\lim_{K\rightarrow\infty}\psi_K(s) = e^{-\lambda(t-s)}$ for almost every $s\in[t_0,t]$.
    Also, for all sufficiently large $K$,
    \begin{equation*}
        \psi_K(s)\|u(s)\|_{L^p}\leq e^{M(t-t_0)/K}\|u(s)\|_{L^p} \leq 2\|u(s)\|_{L^p},
    \end{equation*}
    where $2\|u(s)\|_{L^p}$ is integrable by Hypothesis~\ref{hyp:input}.
    Therefore, by letting $K\rightarrow\infty$ and applying the dominated convergence theorem,
    we obtain \eqref{eqn:iss_bound}.
\end{proof}

\section{Application}
\label{sec:application}

As an application of the established theory, we analyze noise-induced synchronization in a double-well potential from the perspective of stochastic contraction.
We also examine the robustness of synchronization with respect to parameter perturbations based on the $\delta$ISS result.

\subsection{Noise-Induced Contraction and Synchronization}

Consider the stochastic dynamical system
\begin{equation}
\label{eqn:double_well_potential}
    dx = {(-x^3 + x)}dt + dW_t \in \bbR,
\end{equation}
which corresponds to an overdamped Langevin dynamics with a double-well potential $ U(x) = \frac{1}{4}x^4 - \frac{1}{2} x^2$.
In the noise-free case, the double-well structure of the potential prevents the system from being contractive (with respect to any time-independent distance function).
However, the addition of noise can render the system stochastically contracting.
We show this by constructing a Finsler structure satisfying \eqref{eqn:finsler_condition}.

\begin{figure}
    \centering
    \includegraphics[width=0.9\linewidth]{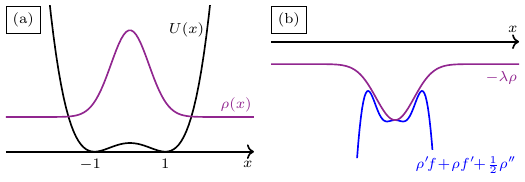}
    \caption{(a) The double-well potential $U(x)$ and the ansatz for $\rho(x)$. (b) The left-hand side and the right-hand side of \eqref{eqn:double_well_condition} with $\lambda=\frac{1}{4}$.}
    \label{fig:ansatz}
\end{figure}

Let us consider a time-independent Finsler structure $F(x,v) = \rho(x)|v|$,
where $\rho:\bbR\rightarrow(0,\infty)$ is a $C^2$ function.
Then, by Theorem~\ref{thm:Lp_contraction}, system~\eqref{eqn:double_well_potential} is $L^1$-contracting if and only if
\begin{equation}
\label{eqn:double_well_condition}
    \rho'(x) f(x) + \rho(x) f'(x) + {\frac{1}{2} \rho''(x)} \leq - \lambda \rho(x)\quad \forall x\in\bbR.
\end{equation}
As discussed in item~\ref{item:noise_induced_contraction} of Section~\ref{sec:Lp_differential_characterization},
condition~\eqref{eqn:double_well_condition} can be satisfied when the noise-induced term $\frac{1}2{}\rho''(x)$ is negative in the deterministically non-contracting region.
In this regard, we adopt an ansatz $\rho(x) = 1+ \frac{5}{2} \exp\!\left(-\frac{7}{4}x^2\right)$,
which is concave around the unstable equilibrium point $x=0$ (see Fig.~\ref{fig:ansatz}(a)).
Indeed, as illustrated in Figure~\ref{fig:ansatz}(b), this choice of $\rho$ satisfies \eqref{eqn:double_well_condition} with $\lambda=1/4$, with equality attained at $x=0$.
Therefore, we can conclude that system~\eqref{eqn:double_well_potential} is $L^1$-contracting in the Finsler distance $\sfd$ induced by $F$ with rate $1/4$.

We now discuss noise-induced synchronization in the double-well potential.
Let $x_1(t)$ and $x_2(t)$ denote the positions of two particles evolving in the same double-well potential, starting from $x_1(0)=\xi_1$ and $x_2(0)=\xi_2$ and driven by a common external noise process.
The $L^1$-contraction property of \eqref{eqn:double_well_potential} provides several quantitative guarantees for their synchronization.
First, \eqref{eqn:Lp_contraction} itself gives an exponentially decaying upper bound on $\bbE[\sfd(x_1(t),x_2(t))]$.
Second, since the Finsler structure is uniform, system~\eqref{eqn:double_well_potential} is $L^1$-UIES in the Euclidean distance with $C=3.5$ by Proposition~\ref{prop:equivalent_distance}.
This yields an exponentially decaying upper bound on $\bbE|x_1(t)-x_2(t)|$.
Third, Proposition~\ref{prop:as_ies_hierarchy} implies that the two particles synchronize exponentially with rate $\lambda$ almost surely.

\begin{figure}
    \centering
    \includegraphics[width=\linewidth]{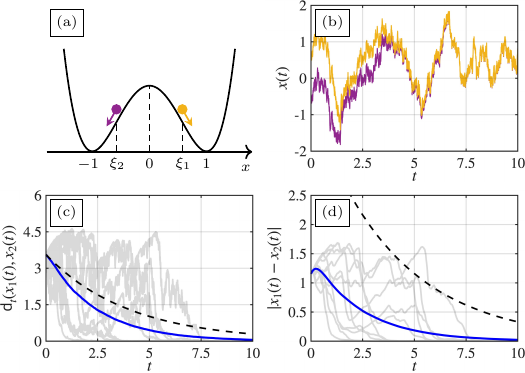}
    \caption{(a) Initial conditions $\xi_1=1/\sqrt{3}$ and $\xi_2=-1/\sqrt{3}$. (b) A pair of sample paths of $x_1(t)$ and $x_2(t)$. (c) (resp.~(d)) The blue solid curve shows a Monte Carlo estimate of $\bbE[\sfd(x_1(t),x_2(t))]$ (resp. $\bbE|x_1(t)-x_2(t)|$) obtained from 5000 samples. The black dashed curve shows the theoretical upper bound $e^{-0.25 t}\sfd(\xi_1,\xi_2)$ (resp. $3.5 e^{-0.25 t}|\xi_1-\xi_2|$), and the gray curves show 10 sample paths of $\sfd(x_1(t),x_2(t))$ (resp. $|x_1(t)-x_2(t)|$).}
    \label{fig:synchronization}
\end{figure}

Numerical simulations support these theoretical results.
As shown in Fig.~\ref{fig:synchronization}(a), we set $\xi_1=1/\sqrt{3}$ and $\xi_2=-1/\sqrt{3}$, which are the points at which the magnitude of the outward drift is maximal.
We then simulated the evolution of two trajectories $(x_1(t))_{t\geq 0}$ and $(x_2(t))_{t\geq0}$ using the Euler-Maruyama scheme.
Fig.~\ref{fig:synchronization}(b) depicts a pair of sample paths, which clearly exhibit synchronization.
Fig.~\ref{fig:synchronization}(c)--(d) depict the Monte Carlo estimates
of $\bbE[\sfd(x_1(t),x_2(t))]$ and $\bbE|x_1(t)-x_2(t)|$, respectively, together with their theoretical upper bound.
The estimate of $\bbE[\sfd(x_1(t),x_2(t))]$ exhibits exponential decay without overshoot, consistent with the theoretically predicted decay rate. Notably, the two curves are nearly tangent at $t=0$, suggesting that the theoretical bound is tight.
In contrast, the estimate of $\bbE|x_1(t)-x_2(t)|$ initially increases due to the repelling drift.

Noise-induced synchronization in a double-well potential was also studied in \cite{crauel1998,flandoli2017} in terms of a random attractor \cite{crauel1994}.
Compared with them, our contraction-based approach additionally establishes exponential synchronization
together with upper bounds on $\bbE[\sfd(x_1(t),x_2(t))]$ and $\bbE|x_1(t)-x_2(t)|$.
These stronger results can facilitate further investigation of the phenomenon, such as synchronization time analysis.
Indeed, using Markov's inequality, one can deduce that
\begin{equation*}
    t\geq T_0 \quad \Rightarrow \quad \bbP(|x_1(t)-x_2(t)| < \varepsilon) \geq 1-\eta,
\end{equation*}
where $T_0\coloneqq {\frac{1}{\lambda}\ln(C|\xi_1-\xi_2|/(\varepsilon\eta))}$, for any $\varepsilon,\eta>0$.
This implies that practical synchronization is achieved with high probability after time $T_0$.

\subsection{Robustness of Synchronization}

\begin{figure}
    \centering
    \includegraphics[width=\linewidth]{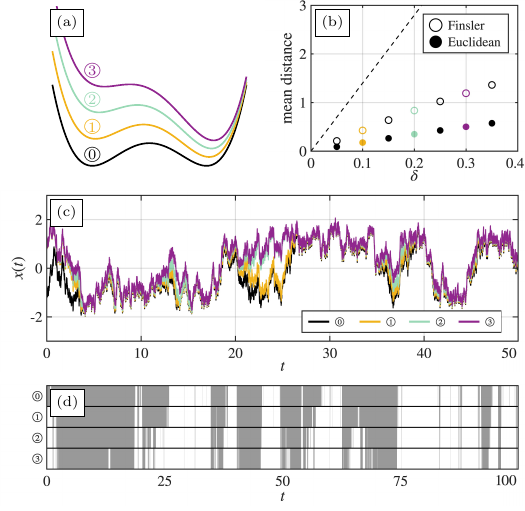}
    \caption{(a) The nominal double-well potential (\circnum{0}) and perturbed, unbalanced double-well potentials (\circnum{1}: $\delta=0.1$, \circnum{2}: $\delta=0.2$, \circnum{3}: $\delta=0.3$).
    (b) Monte Carlo estimates, based on 5000 samples, of $\bbE[\sfd(x_1(100),x_2(100))]$ (open circles) and $\bbE|x_1(100)-x_2(100)|$ (filled circles) for various values of $\delta$. 
    The black dashed lines represent their theoretical upper bounds \eqref{eqn:perturbation_upper_bound} with $t=100$, which are almost indistinguishable.
    (c) A sample path of $x_1(t)$ (\circnum{0}) and sample paths of $x_2(t)$ for $\delta=0.1,0.2,0.3$ (\circnum{1}--\circnum{3}), all based on the same noise realization.
    (d) A raster plot showing the signs of $x_1(t)$ (\circnum{0}) and $x_2(t)$ for $\delta=0.1,0.2,0.3$ (\circnum{1}--\circnum{3}), all based on the same noise realization. White and gray indicate nonnegative and negative states, respectively.
    }
    \label{fig:parameter_perturbation}
\end{figure}

We now investigate the robustness of noise-induced synchronization against a perturbation.
Consider two particles subject to a common noise process
\begin{subequations}
    \begin{align}
        dx_1 &= (-x_1^3+x_1)dt + dW_t,\\
        dx_2 &= (-x_2^3+x_2+\delta)dt + dW_t, \label{eqn:double_well_perturbed}
    \end{align}
\end{subequations}
where the first particle follows the nominal dynamics, and the second follows the dynamics perturbed by $\delta\in\bbR$.
The latter can be interpreted as an overdamped Langevin dynamics in an unbalanced double-well potential (see Fig.~\ref{fig:parameter_perturbation}(a)).
Let $x_1(0)=\xi_1$ and $x_2(0)=\xi_2$ be initial conditions.

By regarding $\delta$ as a constant input in~\eqref{eqn:iss_with_input}, one can observe that Hypothesis~\ref{hyp:iss_regularity} holds with $L_x=1$ and $L_u=1$.
Hence, Theorem~\ref{thm:iss} yields
\begin{subequations}
\label{eqn:perturbation_upper_bound}
\begin{align}
    &\bbE[\sfd(x_1(t),x_2(t))] \leq e^{-\lambda t} \sfd( \xi_1,\xi_2) +\frac{CL_u}{\lambda}|\delta|, \label{eqn:perturbation_finsler_upper_bound}\\
    &\bbE|x_1(t)-x_2(t)| \leq C e^{-\lambda t} |\xi_1- \xi_2| +\frac{CL_u}{\lambda}|\delta| \label{eqn:perturbation_euclidean_upper_bound}
\end{align}
\end{subequations}
with $CL_u/\lambda = 14$.
Consequently, for every $\varepsilon>0$, Markov's inequality gives
\begin{equation}
\label{eqn:perturbation_synchronization_probability}
    \bbP(|x_1(t)-x_2(t)|>\varepsilon) \leq C e^{-\lambda t} \frac{|\xi_1-\xi_2|}{\varepsilon} + \frac{CL_u|\delta|}{\lambda \varepsilon}.
\end{equation}
Thus, the two particles will be practically synchronized with high probability after an initial transient, provided that the parameter perturbation is small.
While the robustness of noise-induced synchronization can also be analyzed using the upper semicontinuity of a random attractor \cite{robinson2002}, the contraction-based approach provides an explicit bound \eqref{eqn:perturbation_synchronization_probability} on the synchronization probability.

We demonstrate the robustness of synchronization through numerical simulations.
The initial conditions are set as $\xi_1 = -1$ and $\xi_2=1$.
For several values of $\delta$, we estimate $\bbE[\sfd( x_1(100),x_2(100))]$ and $\bbE|x_1(100)-x_2(100)|$ via Monte Carlo simulations.
As shown in Fig.~\ref{fig:parameter_perturbation}(b), the results are consistent with the theoretical estimates \eqref{eqn:perturbation_upper_bound}.
Fig.~\ref{fig:parameter_perturbation}(c) shows a sample path of $x_1(t)$ and sample paths of $x_2(t)$ for different values of $\delta$, all based on the same realization of the noise process.
It can be seen that the trajectories remain close for most of the time after an initial transient, while larger values of $\delta$ lead to a brief loss of synchronization around $t\approx 25$.
This effect is more clearly visible in the raster plot in Fig.~\ref{fig:parameter_perturbation}(d), which records the sign of each trajectory over time.
While the overall switching patterns are similar, as $\delta$ increases, the perturbed trajectory tends to remain positive for longer periods.

\section{Conclusion}
\label{sec:conclusion}

This paper studied contraction between trajectories of It{\^o} stochastic differential equations driven by a common white-noise process.
Considering Finsler distances, we derived differential conditions that characterize both $L^p$- and almost sure contraction.
Then, we rigorously proved their sufficiency and necessity under suitable hypotheses.
We also specialized the results to Riemannian and norm-induced distances and established an incremental input-to-state stability result in the common-noise setting.
As an application, we certified noise-induced synchronization in an overdamped Langevin system with a double-well potential using an explicitly constructed Finsler distance, and
quantified its robustness with respect to parameter perturbations.

Several directions remain open. On the theoretical side, it would be of
interest to consider nonsmooth distances, such as those induced by the $1$-norm and $\infty$-norm.
Extending the framework to other noise models, including L{\'e}vy noise as considered in \cite{han2022rnc}, is another natural direction.
Studying the incremental input-to-state stability in a more general setting is also promising.
On the application side, applying the contraction-based approach to
noise-induced synchronization in neuronal oscillators and chaotic systems
remains an open direction.
Common-noise contraction may also be applied to model reduction of singularly
perturbed stochastic systems, paralleling the role of deterministic contraction in establishing approximation guarantees over an infinite time interval \cite{hoppensteadt,kim2015tac,lee2020aut}.

\appendix

\section{Formal Statement of Hypothesis~\ref{hyp:regularity}}
\label{appendix:regularity_details}
Condition~\ref{item:local_Lipschitz_uniformly_in_time} of Hypothesis~\ref{hyp:regularity} is understood as follows: 
For each pair of compact sets $J\subseteq \bbR$ and $K\subseteq \bbR^n$, there exists a constant $C_{J,K}>0$ for which
\begin{equation}
\label{eqn:uniform_local_Lipschitz}
\begin{aligned}
    &\|f(t,x)-f(t,x')\| \leq C_{J,K} \|x-x'\|\\
    &\|\sigma(t,x)-\sigma(t,x')\| \leq C_{J,K} \|x-x'\|
\end{aligned}
\end{equation}
hold for all $t\in J$, $x,x'\in K$.

Condition~\ref{item:local_Holder_uniformly_in_time} is understood as follows: There exists a constant $\alpha>0$ such that, for each pair of compact sets $J\subseteq \bbR$ and $K\subseteq \bbR^n$, there exists a constant $C_{J,K}>0$ for which
\begin{equation}
\label{eqn:uniform_local_Holder}
\begin{aligned}
    &\left\|\frac{\partial f}{\partial x}(t,x) - \frac{\partial f}{\partial x}(t,x')\right\| \leq C_{J,K} \|x-x'\|^\alpha\\
    &
    \left\|\frac{\partial \sigma_j}{\partial x}(t,x) - \frac{\partial \sigma_j}{\partial x}(t,x')\right\| \leq C_{J,K} \|x-x'\|^\alpha
\end{aligned}
\end{equation}
hold for all $t\in J$, $x,x'\in K$, and $j\in\{1,\dots,m\}$.

Although condition~\ref{item:local_Holder_uniformly_in_time} ensures that $f(t,x)$ and $\sigma(t,x)$ are locally Lipschitz in $x$ for each fixed $t\in\bbR$, it does not automatically guarantee the existence of a uniform Lipschitz constant required in \ref{item:local_Lipschitz_uniformly_in_time}.

\section{Change of Exponent in Condition~(\ref{eqn:finsler_condition})}
\label{appendix:change_of_exponent}

Let $q > 0$. Let $z$ and $z'$ stand for any of the variables $t$, $x$, and $v$.
Applying the chain rule for $F_t^p = (F_t^q)^{p/q}$, we obtain
\begin{align*}
    &\frac{\partial F_t^p}{\partial z} = \frac{p}{q}F_t^{p-q}\frac{\partial F_t^q}{\partial z},\\
    &\frac{\partial^2 F_t^p}{\partial z'\partial z} = \frac{p}{q}F_t^{p-q}\frac{\partial^2 F_t^q}{\partial z'\partial z} + \frac{p}{q} \frac{p-q}{q} F_t^{p-2q} \left(\frac{\partial F_t^q}{\partial z'}\right)^\top\frac{\partial F_t^q}{\partial z}.
\end{align*}
Plugging this into \eqref{eqn:finsler_condition} yields
\begin{equation*}
    \frac{p}{q}F_t^{p-q}\left[\frac{\partial F_t^q}{\partial t} + \frac{\partial F_t^q}{\partial x} f + \frac{\partial F_t^q}{\partial v} \frac{\partial f}{\partial x} v\right] + \sum_{j=1}^m \Delta_{j,p} \leq -p\lambda F_t^p,
\end{equation*}
and $\Delta_{j,p}$ can be rewritten as
\begin{align*}
    \Delta_{j,p} \!=\! \frac{p}{q}F_t^{p-q} \Delta_{j,q} \!+\! \frac{p(p\!-\!q)}{2q^2} F_t^{p-2q} \!\left(\frac{\partial F_t^q}{\partial x}\sigma_j \!+\! \frac{\partial F_t^q}{\partial v}\frac{\partial \sigma_j}{\partial x}v\right)^2\!\!.
\end{align*}
Noting that $F_t(x,v)> 0$ when $v\neq 0$, dividing both sides by $(p/q)F_t^{p-q}$ now gives \eqref{eqn:another_finsler_condition}.

\section{Technical Lemmas}
\label{appendix:technical_lemmas}

\begin{lemma}
\label{lem:v_nonzero}
    Suppose that Hypothesis~\ref{hyp:regularity} holds.
    Let $(\xi,\zeta)\in T\bbR^n \setminus 0$, and let $(x(t),v(t))$ be a solution of \eqref{eqn:lifted_sde} with initial condition $(x(t_0),v(t_0))=(\xi,\zeta)$.
    Then, $v(t)\neq 0$ for all $t\geq t_0$ almost surely.
\end{lemma}

\begin{proof}
    Since the coefficients of \eqref{eqn:variational_sde} are Lipschitz in $v$, its solution is unique.
    Meanwhile, it can be seen from \eqref{eqn:matrix_sde} that $(\partial_\xi\phi_{t_0,t}(\xi)\zeta)_{t\geq t_0}$ is a solution of \eqref{eqn:variational_sde} starting from $\zeta$ at $t_0$.
    Thus, we have
    \begin{equation*}
        v(t) = \partial_\xi\phi_{t_0,t}(\xi)\zeta \quad \forall t \geq t_0 \quad \text{a.s.}
    \end{equation*}
    Since $\zeta\neq0$ and $\partial_\xi\phi_{t_0,t}(\xi)$ is nonsingular (Lemma~\ref{lem:stochastic_flow}-(d)), we have $v(t)\neq 0$ for all $t\geq t_0$ almost surely.
\end{proof}

\begin{lemma}
\label{lem:linear_approximation}
    Let $F:T\bbR^n\rightarrow \bbR$ be a Finsler structure and $\sfd$ denote the Finsler distance induced by $F$.
    Suppose that Hypothesis~\ref{hyp:convexity} holds. Then,
    \begin{equation*}
        \lim_{r\downarrow 0} \frac{\sfd (\gamma(0),\gamma(r))}{r} = F(\gamma(0), \dot\gamma(0))
    \end{equation*}
    for any piecewise $C^1$ curve $\gamma:[0,1]\rightarrow\bbR^n$.
\end{lemma}

\begin{proof}
    This is a direct consequence of \cite[Thm.~4.3]{busemann41ams}.
\end{proof}

\begin{lemma}
\label{lem:right_continuity}
    Suppose that Hypothesis~\ref{hyp:regularity} holds. Let $x:\bbR\rightarrow\bbR^n$ be a continuous function. Then, for any $t_0\in\bbR$,
    \begin{align*}
        &\lim_{t \downarrow t_0} f(t,x(t)) = f(t_0,x(t_0)),\\
        &\lim_{t\downarrow t_0} \frac{\partial f}{\partial x}(t,x(t)) = \frac{\partial f}{\partial x}(t_0,x(t_0)).
    \end{align*}
    The same statements hold with $f$ replaced by $\sigma_1,\dots,\sigma_m$.
\end{lemma}

\begin{proof}
    We may assume that $(t,x(t))\in J\times K$,
    where $J\coloneqq [t_0,t_0+1]$ and $K\subseteq \bbR^n$ is the closed unit ball centered at $x(t_0)$.
    Let $C_{J,K}$ be the constant appearing in \eqref{eqn:uniform_local_Lipschitz}.
    Then,
    \begin{align*}
        &\|f(t,x(t))-f(t_0,x(t_0))\|\\
        &\leq \| f(t,x(t))-f(t,x(t_0))\| + \|f(t,x(t_0))-f(t_0,x(t_0))\|\\
        &\leq C_{J,K} \|x(t)-x(t_0)\|+\|f(t,x(t_0))-f(t_0,x(t_0))\|.
    \end{align*}
    Since $x$ is continuous and $f$ is right-continuous in $t$, the right-hand side converges to 0 as $t\downarrow t_0$.
    This proves the first result.
    
    To prove the second result, it suffices to show that $\partial_{x_i} f(t,x(t_0))$ is right-continuous in $t$ for each $i=1,\dots,n$; the rest of the proof can be done analogously. To this end, we define
    \begin{equation*}
        Q_h(t) \coloneqq \frac{f(t,x(t_0)+he_i)-f(t,x(t_0))}{h}
    \end{equation*}
    for $h\in(0,1]$, where $e_i$ denotes the $i$-th standard basis vector of $\bbR^n$.
    Then, for $C_{J,K}$ satisfying \eqref{eqn:uniform_local_Holder}, we have
    \begin{align*}
        &\left\|Q_h(t) - \frac{\partial f}{\partial x_i}(t,x(t_0)) \right\|\\
        & = \frac{1}{h} \left\| \int_0^h\left\{\frac{\partial f}{\partial x_i}(t,x(t_0)+re_i) - \frac{\partial f}{\partial x_i} (t,x(t_0))\right\}dr \right\|\\
        & \leq \frac{1}{h} \int_0^h \left\| \frac{\partial f}{\partial x_i}{(t,x(t_0)+re_i)}- \frac{\partial f}{\partial x_i} (t,x(t_0)) \right\|dr\\
        & \leq \frac{C_{J,K}}{h} \int_0^h r^\alpha dr = \frac{C_{J,K}h^\alpha}{\alpha+1}
    \end{align*}
    for all $t\in J$.
    This in particular implies
    \begin{equation*}
        \lim_{h\rightarrow 0} \sup_{t\in J} \left\|Q_h(t) - \frac{\partial f}{\partial x_i}(t,x(t_0)) \right\| =0,
    \end{equation*}
    i.e., $Q_h(t)$ converges to $\partial_{x_i}f(t,x(t_0))$ uniformly in $t$.
    Since $Q_h(t)$ is right-continuous in $t$, its uniform
    limit $\partial_{x_i}f(t,x(t_0))$ is also right-continuous in $t$.
\end{proof}

\begin{lemma}
\label{lem:probability_vanish}
    Let $\tau$ be the stopping time defined by \eqref{eqn:stopping_time_necessity}. 
    Then, $\bbP(\tau\leq t) = o(t-t_0)$.
\end{lemma}

\begin{proof}
    Using the Markov inequality, one can deduce that
    \begin{align*}
        \bbP(\tau\leq t)
        &= \bbP(\|x_0(t\wedge \tau)-\xi\| \vee \|v_0(t\wedge \tau)-\zeta\| \geq \|\zeta\|/2)\\
        &\leq \frac{2^4}{\|\zeta\|^4} \left({\bbE\|x_0(t\wedge \tau)-\xi\|^4} + \bbE\|v_0(t\wedge \tau)-\zeta\|^4\right).
    \end{align*}
    Noting that
    \begin{align*}
        &x_0(t\wedge \tau)-\xi\\
        &=\int_{t_0}^{t} 1_{\{\tau>s\}} f(s,x_0(s))ds +\int_{t_0}^t 1_{\{\tau>s\}} \sigma(s,x_0(s))dW_s
    \end{align*}
    by the Burkholder-Davis-Gundy inequality \cite[Thm.~19.18]{schilling}, there exists a constant $C>0$ such that
    \begin{equation*}
        \big(\bbE[\|x_0(t\wedge\tau)-\xi\|^4]\big)^{1/4}  \leq C(t-t_0)+C(t-t_0)^{1/2}.
    \end{equation*}
    Note that $1_{\{\tau>s\}}f(s,x_0(s))$ and $1_{\{\tau>s\}} \sigma(s,x_0(s))$ are bounded.
    Therefore, we have
    \begin{equation*}
        \bbE[\|x_0(t\wedge\tau)-\xi\|^4] = o(t-t_0).
    \end{equation*}
    A similar argument shows $\bbE[\|v_0(t\wedge\tau)-\zeta\|^4] = o(t-t_0)$, completing the proof.
\end{proof}

\begin{lemma}
\label{lem:iss_lemma}
    Let $p\geq 1$. Suppose that Hypotheses~\ref{hyp:iss_regularity} and \ref{hyp:input} hold.
    Let $t\in\bbR$, and let $\xi$ be an $\bbR^n$-valued $\calF_{t}$-measurable random variable.
    Let $(x_1(s),x_2(s))_{s\geq t}$ be the solution of \eqref{eqn:iss} with $x_1(t)=x_2(t)=\xi$.
    Then,
    \begin{equation*}
        \|x_1(t+h)-x_2(t+h)\|_{L^p} \leq L_u\int_t^{t+h} e^{M(t+h-s)} \|u(s)\|_{L^p}ds 
    \end{equation*}
    for all $h\geq 0$, where $M\coloneqq L_x + \frac{1+|p-2|}{2}L_x^2$.
\end{lemma}

\begin{proof}
    For simplicity, we denote $z(s)\coloneqq x_1(s)-x_2(s)$. For each $\varepsilon>0$, define $V_\varepsilon(s)\coloneqq \sqrt{\|z(s)\|_2^2+\varepsilon}$.
    Using It{\^o}'s formula, one can deduce that
    \begin{align*}
        &dV_\varepsilon^p(s)\\
        &= \bigg[ pV_\varepsilon^{p-2}(s)z(s)^\top (f(s,x_1,0)-f(s,x_2,u))\\
        & \qquad + \frac{p}{2}V_\varepsilon^{p-2}(s) \|\sigma(s,x_1)-\sigma(s,x_2)\|_F^2\\
        & \qquad + \frac{p(p-2)}{2} V_\varepsilon^{p-4}(s)\|(\sigma(s,x_1)\!-\!\sigma(s,x_2))^\top z(s)\|_2^2 \bigg] ds\\
        & \quad + p V_\varepsilon^{p-2}(s) z(s)^\top (\sigma(s,x_1)-\sigma(s,x_2)) dW_s\\
        &\leq \Big[pMV_\varepsilon^p(s) + pL_u V_\varepsilon^{p-1}(s) \|u(s)\|_2 \Big] ds\\
        & \quad + p V_\varepsilon^{p-2}(s) z(s)^\top (\sigma(s,x_1)-\sigma(s,x_2)) dW_s.
    \end{align*}
    Thus, for $R>0$ and a stopping time $\tau_R \coloneqq \inf\{s\geq t:\|z(s)\|_2 \geq R\}$, by H{\"o}lder's inequality, we have
    \begin{align*}
        &\bbE[V_\varepsilon^p((t+h)\wedge \tau_R)]\\
        &\leq  \varepsilon^{p/2} + pM\int_t^{t+h} \bbE[1_{\{\tau_R > s\}}V_\varepsilon^p(s)]ds\\
        &\quad + pL_u \int_t^{t+h} \bbE[1_{\{\tau_R>s\}} V_\varepsilon^{p-1}(s) \|u(s)\|_2] ds\\
        &\leq \varepsilon^{p/2} + pM\int_t^{t+h} \bbE[V_\varepsilon^p(s\wedge \tau_R)]ds\\
        &\quad + pL_u \int_t^{t+h} (\bbE[V_\varepsilon^p(s\wedge \tau_R)])^{(p-1)/p} \|u(s)\|_{L^p}ds\\
        &\eqqcolon H_{\varepsilon,R}^p(t+h)
    \end{align*}
    for all $h\geq 0$.
    We claim that
    \begin{equation}
    \label{eqn:claim_H}
        H_{\varepsilon,R}(t+h) \leq \sqrt{\varepsilon} e^{Mh} + L_u \int_t^{t+h} e^{M(t+h-s)}\|u(s)\|_{L^p}ds.
    \end{equation}
    By definition, $H_{\varepsilon,R}^p(t+h)$ is differentiable almost everywhere, and the derivative satisfies
    \begin{align*}
        &\frac{d}{ds} H_{\varepsilon,R}^p(s)\\
        &= pM \bbE[V_\varepsilon^p(s\wedge \tau_R)] + pL_u (\bbE[V_\varepsilon^p(s\wedge \tau_R)])^{(p-1)/p} \|u(s)\|_{L^p}\\
        & \leq pM H_{\varepsilon,R}^p(s) + pL_u H_{\varepsilon,R}^{p-1}(s) \|u(s)\|_{L^p}
    \end{align*}
    almost everywhere. Since $H_{\varepsilon,R}^p(s)\geq \varepsilon^{p/2} > 0$, by the chain rule,
    \begin{equation*}
        \frac{d}{ds} H_{\varepsilon,R}(s) \leq MH_{\varepsilon,R}(s) + L_u \|u(s)\|_{L^p}
    \end{equation*}
    almost everywhere.
    Therefore, we have
    \begin{equation*}
        \frac{d}{ds} e^{M(t+h-s)} H_{\varepsilon,R}(s) \leq L_u e^{M(t+h-s)}\|u(s)\|_{L^p}
    \end{equation*}
    almost everywhere, and integrating both sides yields the desired inequality \eqref{eqn:claim_H}.
    
    Due to the existence of global solutions, $\lim_{R\rightarrow\infty} \tau_R = \infty$ almost surely.
    Also, $\lim_{\varepsilon\downarrow 0}V_\varepsilon(s) = \|z(s)\|_2$.
    Thus, by Fatou's lemma, we can conclude that
    \begin{align*}
        \| z(t+h)\|_{L^p}^p &\leq \liminf_{\varepsilon\rightarrow 0} \liminf_{R\rightarrow\infty} \bbE[V_\varepsilon^p((t + h)\wedge \tau_R)]\\
        &\leq \liminf_{\varepsilon\rightarrow 0} \liminf_{R\rightarrow\infty}  H_{\varepsilon,R}^p(t+h).
    \end{align*}
    Plugging in \eqref{eqn:claim_H} now completes the proof.
\end{proof}

\section*{Acknowledgements}
The authors thank Rodolphe Sepulchre and Dongjun Wu for valuable discussions.

\section*{Declaration of generative AI use}

The authors used GPT-6.1 Sol to assist with reviewing and proofreading the manuscript.
During this process, the need for strong completeness (rather than weak completeness) in Section 2.1 was identified.
The authors independently verified this point and revised the manuscript accordingly.

\bibliographystyle{unsrt}
\bibliography{autosam}

\scriptsize

\end{document}